\documentclass[fleqn,usenatbib]{mnras}

\usepackage{newtxtext,newtxmath}

\usepackage[T1]{fontenc}

\DeclareRobustCommand{\VAN}[3]{#2}
\let\VANthebibliography\thebibliography
\def\thebibliography{\DeclareRobustCommand{\VAN}[3]{##3}\VANthebibliography}

\usepackage{graphicx}	
\usepackage{amsmath}	
\usepackage{lscape}

\newcommand{\teff}{T$_{\text{eff}}$\,}
\newcommand{\logg}{log\,$g$}
\newcommand{\msun}{M$_\odot$}
\newcommand{\rsun}{R$_\odot$}
\newcommand{\vsini}{$v\sin{i}$}
\newcommand{\mearth}{M$_\oplus$}
\newcommand{\rearth}{R$_\oplus$}
\newcommand{\plusmin}{~$\pm$~}

\newcommand{\kms}{$km s^{-1}$}
\newcommand{\ms}{$m s^{-1}$}
\newcommand\nodata{ ~$\cdots$~ }

\title[Improving Exoplanetary Science with GALAH]{The GALAH Survey: Improving our understanding of confirmed and candidate planetary systems with large stellar surveys}

\author[J.~T.~Clark et al.]{Jake~T.~Clark,$^{1}$\thanks{E-mail: jake.clark@usq.edu.au}
Duncan~J.~Wright,$^{1}$
Robert~A.~Wittenmyer,$^{1}$
Jonathan~Horner,$^{1}$
\newauthor
Natalie~R.~Hinkel,$^{2}$
Mathieu~Clert{\'e},$^{1}$
Brad~D.~Carter,$^{1}$
Sven~Buder,$^{3,4}$
\newauthor
Michael~R.~Hayden,$^{5,4}$
Joss~Bland-Hawthorn,$^{5,4}$
Andrew~R.~Casey,$^{6,7}$
\newauthor
Gayandhi~M.~De~Silva,$^{8,9}$
Valentina~{D'Orazi},$^{10}$
Ken~C.~Freeman,$^{3}$
Janez~Kos,$^{11}$
\newauthor
Geraint~F.~Lewis,$^{5}$
Jane~Lin,$^{3,4}$
Karin~Lind,$^{12,15}$
Sarah~L.~Martell,$^{14,4}$
\newauthor
Katharine~J.~Schlesinger,$^{3}$
Sanjib~Sharma,$^{5,4}$
Jeffrey~D.~Simpson,$^{14,4}$
Dennis~Stello,$^{5,4,14}$
\newauthor
Daniel~B.~Zucker,$^{13,9}$
Toma\v{z}~Zwitter,$^{11}$
Ulisse Munari,$^{10}$
Thomas Nordlander,$^{3,4}$
\\
$^{1}$Centre for Astrophysics, University of Southern Queensland, West Street, Toowoomba, QLD 4350, Australia\\
$^{2}$Space Science and Engineering Division, Southwest Research Institute, San Antonio, TX 78238, USA\\
$^{3}$Research School of Astronomy \& Astrophysics, Australian National University, ACT 2611, Australia\\
$^{4}$Centre of Excellence for Astrophysics in Three Dimensions (ASTRO-3D), Australia\\
$^{5}$Sydney Institute for Astronomy, School of Physics, A28, The University of Sydney, NSW 2006, Australia\\
$^{6}$Monash Centre for Astrophysics, Monash University, Australia\\
$^{7}$School of Physics and Astronomy, Monash University, Australia\\
$^{8}$Australian Astronomical Optics, Faculty of Science and Engineering, Macquarie University, Macquarie Park, NSW 2113, Australia\\
$^{9}$Macquarie University Research Centre for Astronomy, Astrophysics \& Astrophotonics, Sydney, NSW 2109, Australia\\
$^{10}$Istituto Nazionale di Astrofisica, Osservatorio Astronomico di Padova, vicolo dell'Osservatorio 5, 35122, Padova, Italy\\
$^{11}$Faculty of Mathematics and Physics, University of Ljubljana, Jadranska 19, 1000 Ljubljana, Slovenia\\
$^{12}$Department of Astronomy, Stockholm University, AlbaNova University Centre, SE-106 91 Stockholm, Sweden\\
$^{13}$Max Planck Institute for Astronomy, K\:{o}nigstuhl 17, 69117 Heidelberg, Germany\\
$^{14}$School of Physics, University of New South Wales -- Sydney, Sydney 2052, Australia\\
$^{15}$Department of Physics and Astronomy, Macquarie University, Sydney, NSW 2109, Australia\\
}

\date{Accepted XXX. Received YYY; in original form ZZZ}

\pubyear{2021}

\begin{document}
\label{firstpage}
\pagerange{\pageref{firstpage}--\pageref{lastpage}}
\maketitle

\begin{abstract}

\noindent Pioneering photometric, astrometric, and spectroscopic surveys are helping exoplanetary scientists better constrain the fundamental properties of stars within our galaxy, and the planets these stars host. In this study, we use the third data release from the stellar spectroscopic GALAH Survey, coupled with astrometric data of eDR3 from the \textit{Gaia} satellite, and other data from NASA's Exoplanet Archive, to refine our understanding of 279 confirmed and candidate exoplanet host stars and their exoplanets. This homogenously analysed data set comprises 105 confirmed exoplanets, along with 146 K2 candidates, 95 TESS Objects of Interest (TOIs) and 52 Community TOIs (CTOIs). Our analysis significantly shifts several previously (unknown) planet parameters while decreasing the uncertainties for others; Our radius estimates suggest that 35 planet candidates are more likely brown dwarfs or stellar companions due to their new radius values. We are able to refine the radii and masses of WASP-47 e, K2-106 b, and CoRoT-7 b to their most precise values yet, to less than 2.3\% and 8.5\% respectively. We also use stellar rotational values from GALAH to show that most planet candidates will have mass measurements that will be tough to obtain with current ground-based spectrographs. With GALAH's chemical abundances, we show through chemo-kinematics that there are five planet-hosts that are associated with the galaxy's thick disc, including  NGTS-4, K2-183 and K2-337. Finally, we show there is no statistical difference between the chemical properties of hot Neptune and hot rocky exoplanet hosts, with the possibility that short-period rocky worlds might be the remnant cores of hotter, gaseous worlds.


\end{abstract}




\section{Introduction}

Over the past three decades, new discoveries have moved the study of alien worlds from science fiction to science fact. We now know that planets are ubiquitous - that virtually every star in the sky is accompanied by a retinue of exoplanets \citep{Batalha13,Dressing13,Dressing15,Hsu19,Kunimoto20,Yang20}.
One of the great revelations of the first thirty years of the exoplanet era is the diversity exhibited by the physical properties of the planets we have discovered. We have found massive planets with extremely inflated, tenuous atmospheres, resulting in bulk densities lower than that of cotton candy \citep{Rowe14,Dragomir19,Deleuil08,Siverd12,Bailes11}, and others whose density far exceeds that of the planets in the Solar system \citep{Steffen2013,Masuda14,Zhou17,Hartman19,Rowe2014,Lam17,Hartman16}.
Such studies have even revealed systems with similar-sized planets of very disparate mass -- such as the Kepler-107 system; Kepler-107~b and c both have radii of $\sim$1.5\rearth, but while b has an Earth-like density of 5.3 g\,cm$^{-3}$, c has a density at least twice as large, 12.6 g\,cm$^{-3}$; \citealt{bonomo19}). 

In the coming years, our ability to determine the true mass and radius of newly discovered exoplanets will become even more important, as the search for exoplanets becomes a dedicated effort to find planets that could be considered truly `Earth-like'. Those planets will require significant effort to measure their atmospheres in an attempt to look for potential biomarkers. As a result, it will be vitally important to be able to work out which of those planets are most likely to be truly Earth-like \citep[see e.g.][, and references therein]{WhichExoEarth}. 

One of the key pieces of information for that selection process will be precise estimates of the mass and density of the planets in question -- data that can be used to estimate the physical makeup and internal structure of those planets. However, the precision with which we can determine both the mass and the radius of a given planet is limited by the precision with which we can categorise that planet's host star. The less precisely we know the mass or radius of that star, the less accurate will be the parameters we derive for the planets orbiting the star. Our knowledge of the worlds we discover is limited by our knowledge of their host stars.

The problem has become even more pronounced since the launch of \textit{TESS}, which has already delivered a plethora of new discoveries \citep[e.g.][]{Huang18,TOI120,Vanderburg19,AUMicB,TOI677b,TOI257,Rodriguez2021} along with candidate systems waiting for their planetary status to be confirmed, such as \textit{TESS} Object of Interest (TOIs) and Community TOIs (CTOIs). Currently, the median error for TOI planetary radii is in excess of 10\%. Fortunately, vast galactic archaeology surveys are now gathering high-quality spectroscopic data on hundreds of thousands of stars in the local solar neighbourhood, as well as stars within the galaxy's thick and thin disks \citep{LAMOST,RAVE,APOGEE,GAIAESO,DeSilva15}. The GALAH (GALactic Archeology with HERMES) survey \citep{DeSilva15,GALAHDR1,GALAHDR2,GALAHDR3} is one such program, with the aim of gathering high-quality spectra for up to one million nearby stars. In GALAH's latest data release, GALAH DR3 \citep[DR3;][]{GALAHDR3}, the GALAH team provide the results of their observations of 588,571 stars -- including elemental abundances for up to 27 different elements. The resulting dataset is a treasure-trove of information that is of vital importance to the exoplanet community. In particular, the GALAH stellar abundances can assist in determining the compositions of the planets these stars might host \citep{Bitsch20,GALAHTESS,Unterborn2018,Dorn2017,Dorn2015,Rogers10}, and even the types of planets the stars could potentially host \citep{Hinkel2019}. At the same time, the \textit{Gaia} spacecraft is taking incredibly precise distance measurements of the stars being surveyed by GALAH (surveying up to $\sim$ 2\% of all stars in the Galaxy).

In this work, we use data from the GALAH DR3 and \textit{Gaia} EDR3 to refine the stellar parameters for a total of 273 stars in the \textit{TESS} Input Catalog (TIC). We show how GALAH data can greatly improve the precision of our characterisation of potential planet hosting stars.
Section 2 determines the stellar parameters for our sample and comparing them to our surveys. In section 3, we use these stellar parameters to then recharacterise confirmed and candidate planetary systems. We discuss our results in Section 4, identifying false positives, using our stellar \vsini~values to determine how difficult it will be to determine mass measurements for our candidates, discuss the ultra-short period planets within our sample, confirm thick-disc host stars, compare the chemical abundances of hot Neptune and hot rocky worlds, and assess the radius valley and super-Earth desert of our sample. Finally we give our conclusions in Section 5.

\begin{figure*}
 \centering
	\includegraphics[width=0.9\columnwidth]{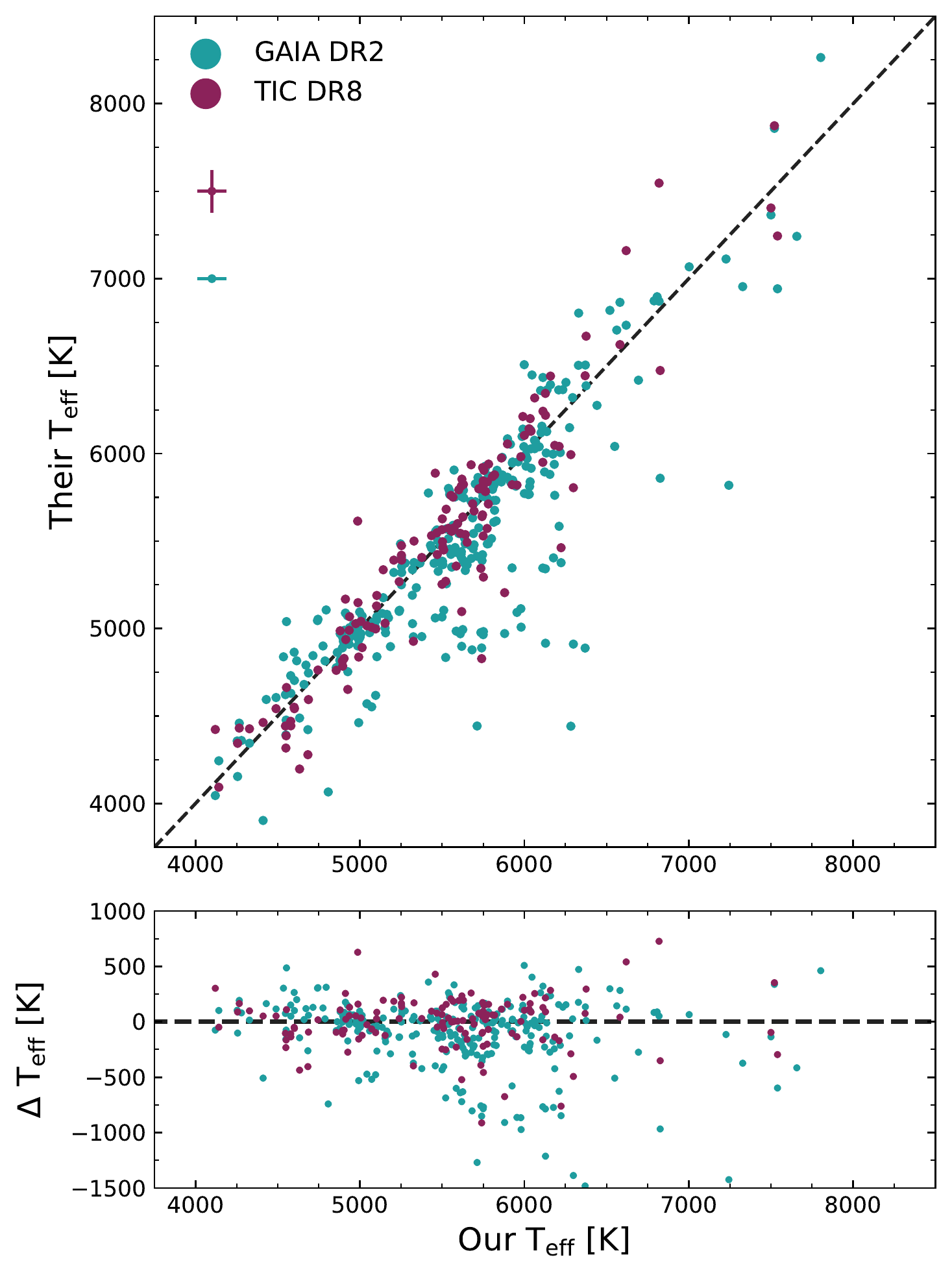}
	\includegraphics[width=1.05\columnwidth]{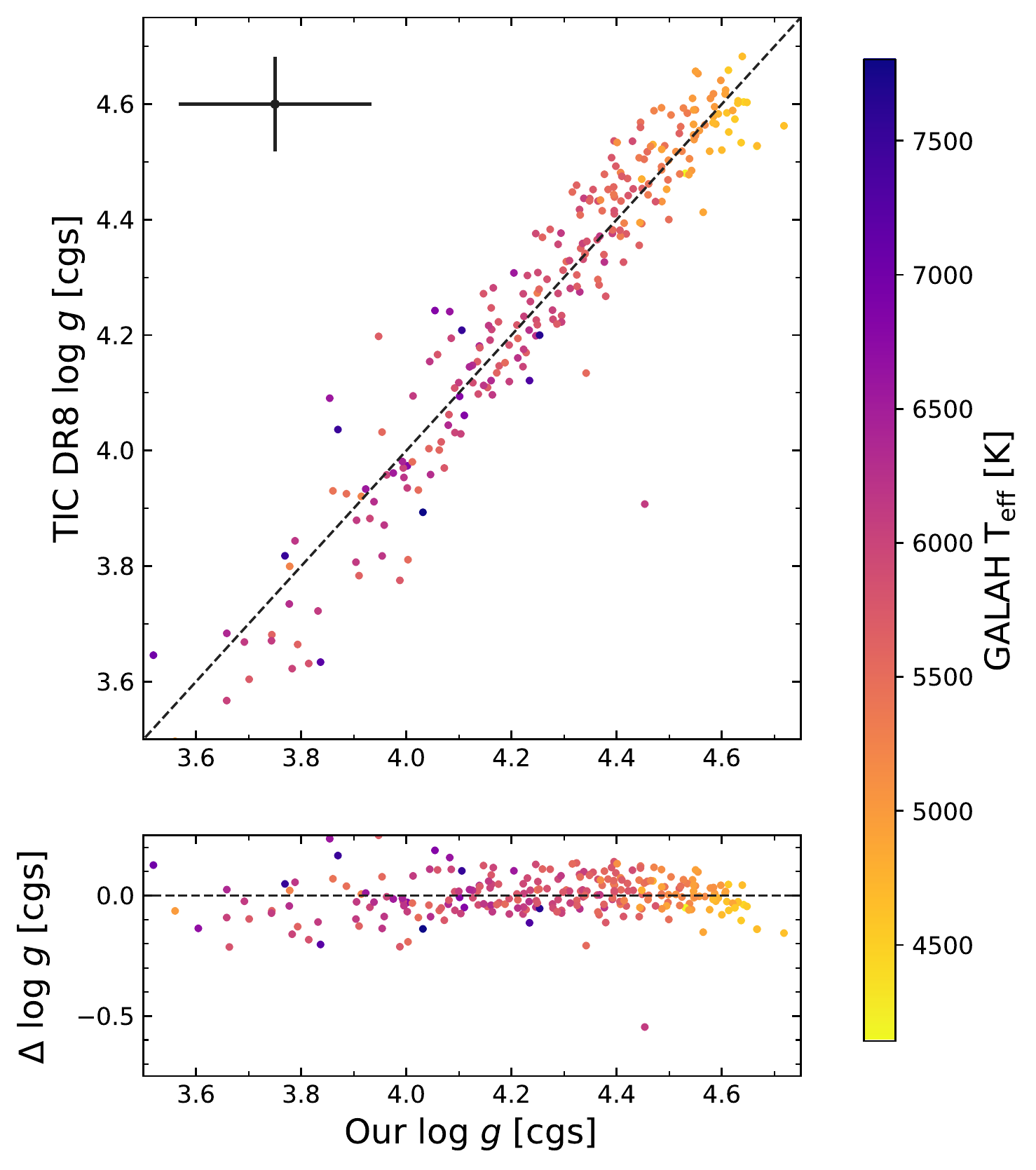}
 \caption{\textbf{Left:} GALAH DR3's \teff results are plotted against the \teff measurements found within the \textit{Gaia} DR2 (blue) and TIC (purple) catalogs respectively. An equality line (grey dashed) is plotted for comparison. Only stars within the TIC that were flagged as obtaining their effective temperature values from photometric surveys are plotted on this graph. A median GALAH-TIC and GALAH-\textit{Gaia} \teff error bars are found in the figure's top left (purple and blue crosses respectively). Because the \textit{Gaia} DR2 catalog currently does not produce errors for their \teff values, there is no median y-error bar for \textit{Gaia} in the plot. The differences in our results are plotted in the bottom plot with $\Delta$X = X\textsubscript{Other} - X\textsubscript{GALAH}. \textbf{Right:} GALAH's \logg~values compared to the TICs with each star colour-coded by their \teff values. As with the \teff figure to the left, an equality line (grey-dashed line) is shown along with the median TESS and GALAH \logg~error bar shown in the top left (grey cross). \logg~values for giant stars are not derived in the TIC and are thus not compared with GALAH's values within this figure.}\label{fig:tefflogg}
\end{figure*}

\section{Determining Stellar Parameters}\label{sec:stellarparam}

\subsection{Combining Input Data}

All three GALAH Data Releases \citep[DR1; DR2; DR3;][respectively]{GALAHDR1,GALAHDR2,GALAHDR3} contain physical and chemical parameters for stars observed on the 3.9~m Anglo-Australian Telescope (AAT), situated on Gamilaraay land in New South Wales, Australia. There are 392 science fibres attached to the two degree-field prime focus top-end \citep[2dF;][]{AAT2dF} that feed into the High Efficiency and Resolution Multi Element Spectrograph (HERMES) \citep{HERMES15}, delivering high resolution (R $\approx$28,000) spectra in four distinct wavelength arms covering 471.3-490.3\,nm, 564.8-587.3\,nm, 647.8-673.7\,nm and 758.5-788.7\,nm respectively. Systematic and atmospheric effects are corrected for each spectrum and then continuum normalised with detailed physical parameters including \teff, \logg, [M/H] and individual abundances derived from each stellar spectrum using 1D stellar atmospheric models via the Spectroscopy Made Easy \citep[or SME, ][]{SMEold} package. 

\subsubsection{GALAH DR3}

GALAH DR3 is slightly different from its previous two counterparts (GALAH DR2; \citealt{GALAHDR2}, and DR1; \citealt{GALAHDR1}) in a few ways. Firstly, GALAH DR3 includes other stellar surveys, such as \textit{TESS}-HERMES \citep{TESSHERMES} and K2-HERMES \citep{K2HERMESours,K2HERMESsanjib}, that have also used the AAT and HERMES instrument. This creates a catalog with more coverage across the ecliptic and southern ecliptic pole. It also creates a catalog that includes stars specifically observed for exoplanet detection and characterisation.

Secondly, GALAH DR3 solely uses SME to derive stellar parameters for all $\sim$600,000 stars, whereas SME was only used in a subset of stars within DR1 and DR2 with these results then propagated through the rest of the catalog thanks to \textsc{The Cannon} \citep{thecannon}. Quality flags have been determined for each star, encoded as a bitmask with flags raised indicating various problems with the analysis. For our analysis, we only include stars within our analysis that had an SME flag \texttt{flag\char`_sp} = 0, and also required the Fe abundance flag \texttt{flag\char`_X\char`_fe} = 0 in the GALAH DR3 release.

\begin{figure}
 \centering
	\includegraphics[width=\columnwidth]{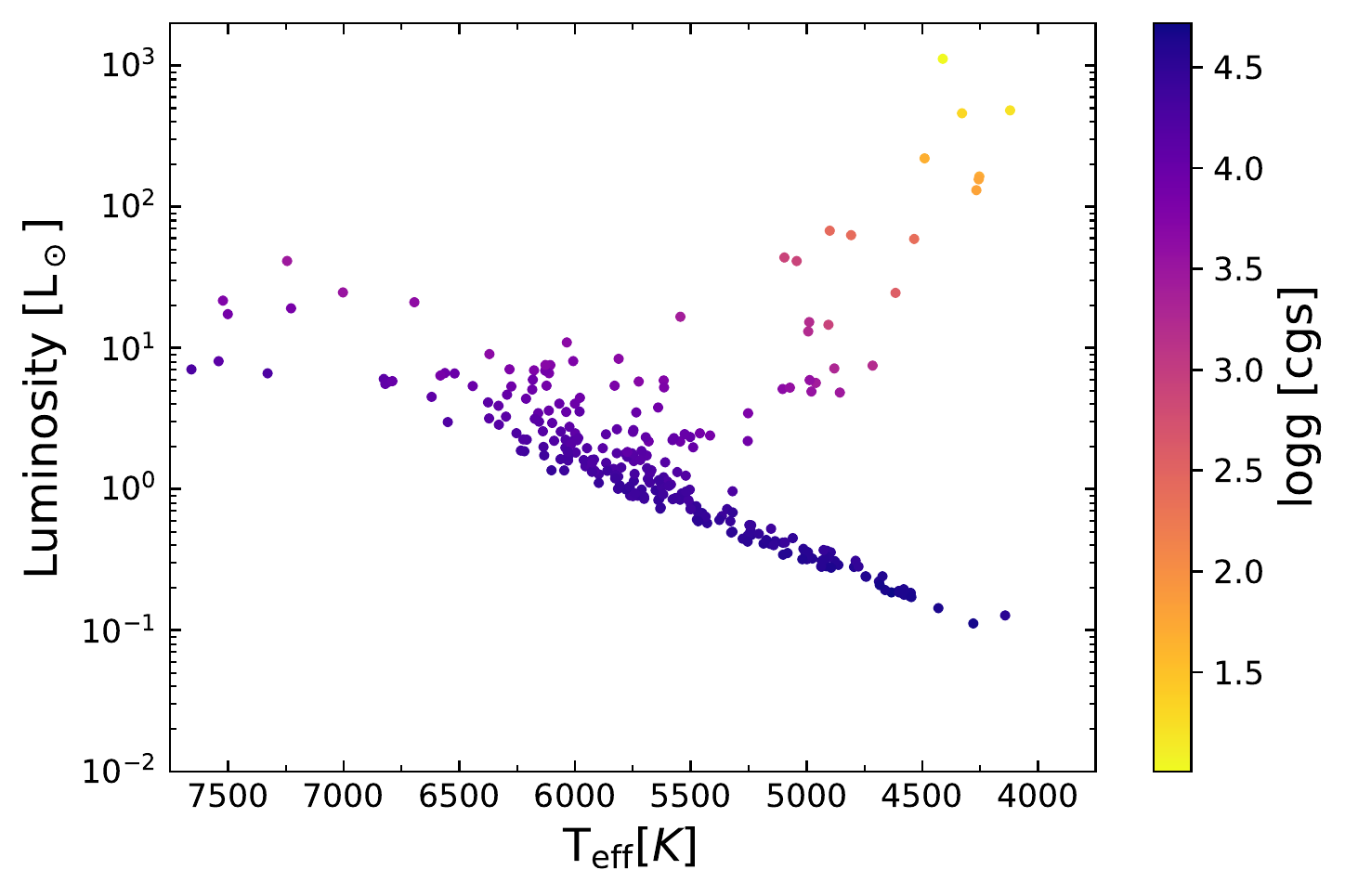}
 \caption{This Hertzsprung-Russell diagram of our GALAH DR3 \teff and \textsc{isochrones} luminosity values colour-coded by GALAH DR3's \logg values, shows that the vast majority of planet-hosting or candidate-system stars lie on the main sequence. Of our 280 stars, only 21 are giants (\teff $<$ 5500 and \logg $<$ 3.5)}\label{fig:hertz}
\end{figure}

\subsubsection{Cross-matching with other stellar and planetary catalogues}

For our analysis, we firstly cross-matched GALAH DR3 with the TIC DR8 \citep{TIC_CTL}, \textit{Gaia} DR2 \citep{gaiadeearetwo} and \textit{Gaia} EDR3 \citep{GAIAEDR3} using the TOPCAT \citep{TOPCAT} tool, cross-matching both catalogs with a position tolerance of $\pm$1\arcsec. Because we are interested in re-characterising confirmed or candidate exoplanetary systems, we then cross-matched GALAH DR3 with four different exoplanet and exoplanet candidate catalogs. These four catalogs include the NASA's Exoplanet Archive (NEA), and exoplanetary candidates from the K2\footnote{\url{https://exoplanetarchive.ipac.caltech.edu/}; accessed 23 November 2021.} and TESS missions\footnote{\url{https://exofop.ipac.caltech.edu/}; accessed 23 November 2021.}. We cross-matched the NEA, and TESS TOI and CTOI catalogs by comparing their TIC identifiers with those found within our catalog, and cross-matched the K2 candidate list with our catalog by comparing 2MASS identifiers. We only accepted a K2 cross-match if the planet candidate was flagged as a ``CANDIDATE" within the K2 Candidate Catalog, rejecting those which are already confirmed exoplanets, which would be picked up through the Exoplanet catalog cross-match, and already known false-positives. We have included all TOIs and CTOIs, regardless of their disposition currently described by the \textit{TESS} Follow-up Observing Program Working Group (TFOPWG) within the TOI and CTOI catalogs. From this cross-matching, we have identified 280 stars within GALAH DR3 that either host confirmed or candidate exoplanets.

\begin{figure*}
 \centering
	\includegraphics[width=\columnwidth]{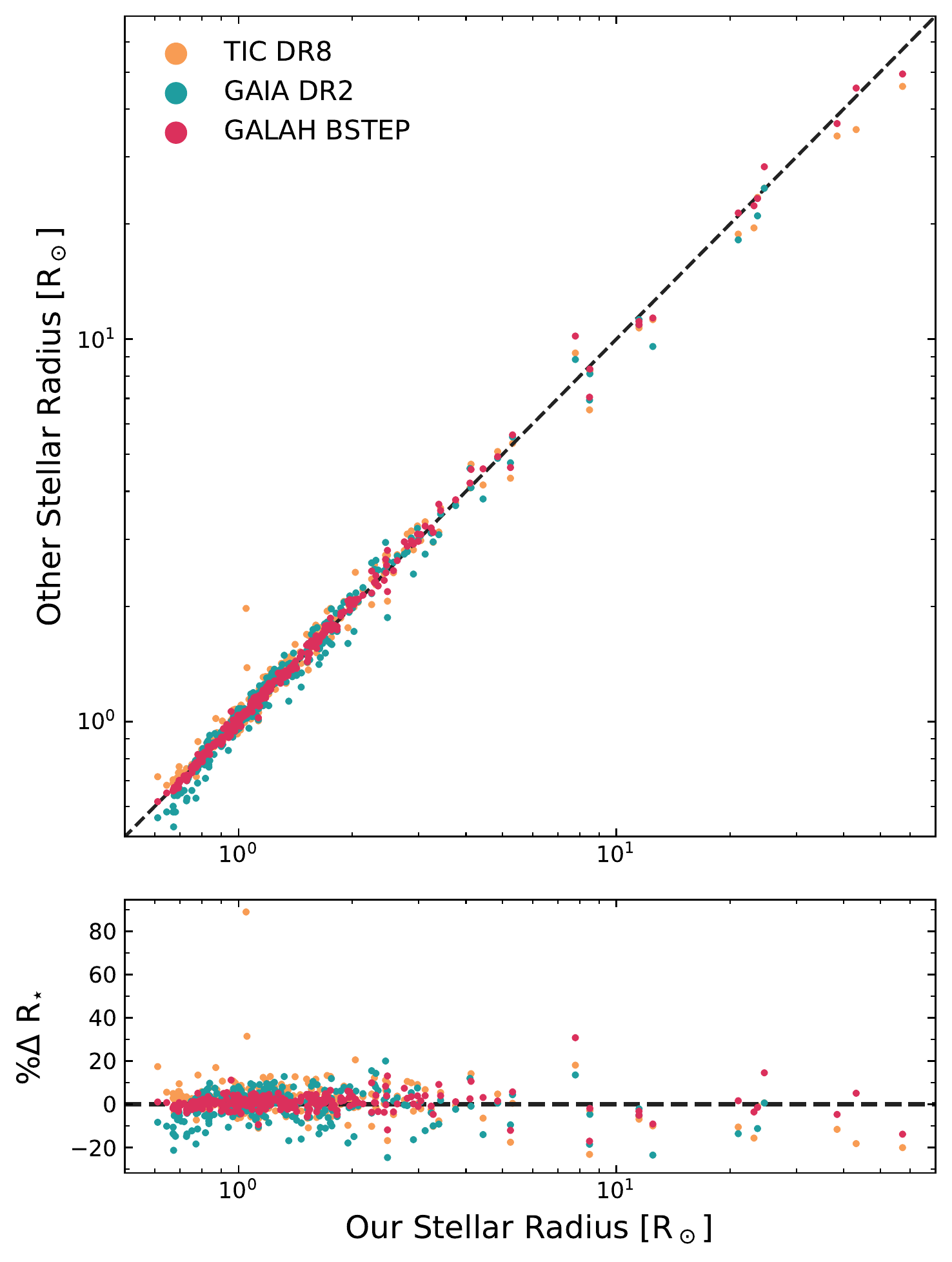}
	\includegraphics[width=\columnwidth]{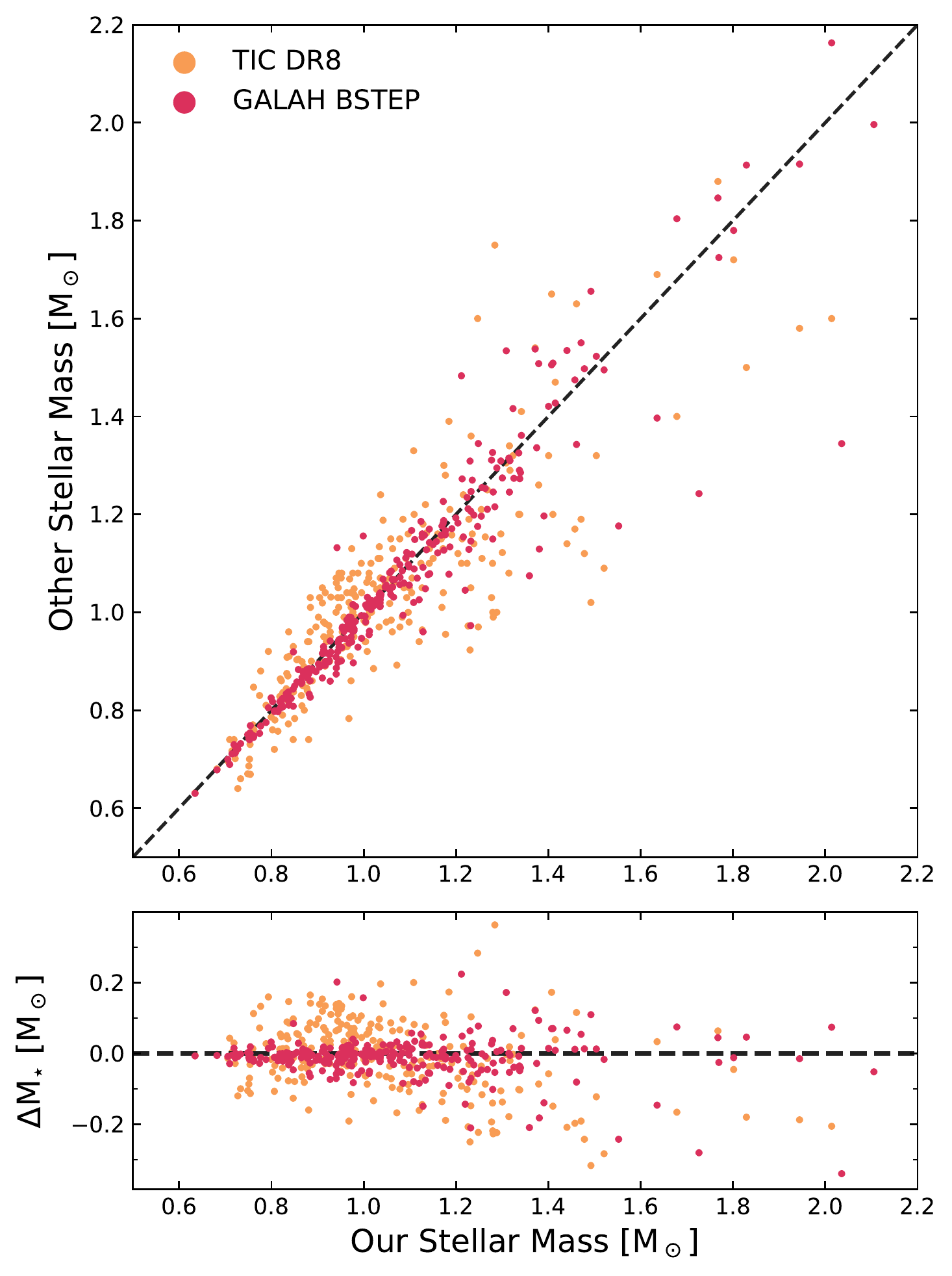}
 \caption{\textbf{Left:} Our stellar radius values are compared to other surveys and methodologies to confirm the validity of our values. Radius values derived from \textsc{isochrones} are compared with values found within the TIC (yellow), \textit{Gaia} DR2 (blue) and GALAH's BSTEP VAC (red). Error bars are suppressed for clarity. An equality line in both plots is depicted in a grey dashed line. \textbf{Right:} Our stellar mass values are compared against those that are found within the TIC (yellow) and GALAH's BSTEP VAC (red). 1-$\sigma$ median error bars for each comparison can be found in the top-right corner of the figure.}\label{fig:starradmass}
\end{figure*}

\subsection{Stellar Masses, Radii, Luminosities, Ages via \textsc{Isochrones}}

To derive the physical parameters of these host stars from the GALAH catalog, we use a similar approach to \citet{GALAHTESS}, implementing the Python package \textsc{isochrones} \citep{isochrones}. \textsc{isochrones} derives physical parameters from observed stellar parameters using the stellar evolution grid, MESA Isochrones \& Stellar Tracks \citep[MIST;][]{MIST}. For our analysis, we used the star's effective temperature (\teff), surface gravity (log\,$g$) and colour excess (E(B-V)) from GALAH, Johnson-Morgan ($V$; taken from the TIC), 2MASS ($J$, $H$, $K_s$) and \textit{Gaia} EDR3 ($G$, $G_{RP}$, $G_{BP}$) photometric magnitudes\footnote{We note that the transmission curves for \textit{Gaia}'s photometric bands are different from DR2 to EDR3. Currently, \textsc{isochrones} only handles \textit{Gaia} DR2 photometric bands. We ran tests during this analysis for the impact of using the different photometric values from DR2 to EDR3 and found changes to the astrophysical properties of the stars well within their 1-sigma error bars.}, along with parallax values obtained by \textit{Gaia} EDR3 \citep{GAIAEDR3} where available as input observables for \textsc{isochrones}. We incorporated the associated errors for each measurement from their respective catalogs but inflated the associated errors in the photometric bands to 0.05 mag. By inflating the magnitude errors rather than including their associated errors, we found the stars to better converge on more realistic results, with \textsc{isochrones} producing better estimates on the error of the derived parameters. We also include the star's global metallicity, [M/H], derived by GALAH's [Fe/H] and [$\alpha$/Fe] abundances formulated by \citet{Salaris06} as:

\begin{equation}\label{eq:fehmh}
\mathrm{[M/H] = [Fe/H]} + \log_{10}{\big(0.694f_\alpha + 0.306\big)}\,\,,
\end{equation}

\noindent where $f_\alpha$ is the $\alpha$-element enhancement factor given by $f_\alpha = 10^{\left[\frac{\alpha}{Fe}\right]}$. When the model reaches convergence, the median and corresponding 1-$\sigma$ uncertainties for stellar mass, radius, density, age, and equivalent evolution phase values are calculated from their respective posterior distributions. A star's stellar luminosity is then calculated through the Stefan-Boltzmann relationship, with these resulting luminosities used to derive the habitable zone boundaries for each star, as formulated by \citet{HZcalc}. We include all five HZ boundaries within \citet{HZcalc} including Recent Venus, Runaway Greenhouse, Moist Greenhouse, Maximum Greenhouse and Early Mars. Rotational, radial and microturbulence velocities from GALAH DR3 have also been included in our stellar parameter table (Table \ref{tab:stellarparams}), to assist ground-based radial velocity teams to better prioritise follow-up targets, including K2, TOI and CTOI candidates.

\begin{table}
\caption{Median and 1$\sigma$ values of our exoplanet/candidate-hosting population}
\begin{center}
\begin{tabular}{rl}
\hline
Quantity & $\mu \pm \sigma$ \\
\hline
\teff~[K]& 5633 $\pm$ 88\\
\logg~[cgs]& 4.28 $\pm$ 0.19 \\
{[Fe/H]} [dex] & -0.01 $\pm$ 0.08 \\
{[$\alpha$/Fe]} [dex] & 0.02 $\pm$ 0.04 \\
{[M/H]} [dex] & 0.03 $\pm$ 0.10 \\
R$_\star$ [R$_\odot$]& 1.18 $\pm$ 0.03 \\
M$_\star$ [M$_\odot$]& 1.02 $\pm$ 0.04 \\
L$_\star$ [L$_\odot$]& 1.49 $\pm$ 0.11 \\
Age [Gyr] & 4.98 $\pm$ 2.90 \\
\hline    
\end{tabular}
\end{center}
\label{tab:medstellar}
\end{table}

\subsection{Comparison of Stellar Parameter estimates}

Because our methodology uses cross-matched data with \textit{Gaia} DR2 and TIC DR8, we can use stellar parameter values from these catalogs to determine any biases from our own data. The median value and median 1-$\sigma$ errors for our stellar parameters can be found in Table\ref{tab:medstellar} with the full stellar parameter catalog being found in Table \ref{tab:stellarparams}. Figure \ref{fig:tefflogg} shows the comparison of GALAH DR3 \logg~and \teff~values with those determined by \textit{Gaia} and the TIC.

\subsubsection{Stellar \teff}

There is an overall good agreement between the catalogs, with a mean difference between Teff values from GALAH and TIC of $-11^{+121}_{-127}$~K, and \textit{Gaia} of $32^{+281}_{-163}$~K, with an RMS difference of 10 and 21 K, respectively. Just as \citet{KHU20} and \cite{GALAHTESS} showed in their work, we also find a horizontal structure within our \textit{Gaia} \teff~comparison near 5000~K. This further suggests that \textit{Gaia} \teff~values preferentially converge onto some \teff values over others. For our comparison of TIC \teff~values, we only compare GALAH's values to those found with the TIC that have been derived photometrically and not spectroscopically. This is to avoid the potential comparison of GALAH DR2 values, as the TIC incorporated GALAH DR2's \teff~measurements into their catalog. Our median planet-hosting temperature of 5514~K is slightly cooler than candidate-hosts with a median temperature of 5698~K.

\subsubsection{Stellar \logg}

Because the \textit{Gaia} catalog contains no \logg~values for their stellar catalog, we only compare our \logg~results to those found within the TIC. In its current iteration, the TIC has \logg~values for only dwarf stars and not giants, to fit within the science requirements of the \textit{TESS} mission. Figure \ref{fig:tefflogg} also shows the comparison with our dwarf \logg~values to those found in the TIC. There is a strong overall agreement with our \logg~values and those found within the TIC with a bias and RMS of ${0.00}^{+0.06}_{-0.08}$ and 0.01 dex respectively. Only one star seemed to be a significant outlier with the CTOI-host star TIC 179582003 having a \logg~of 4.45$\pm$0.19 dex compared to a slightly lower surface gravity found within the TIC of 3.91$\pm$0.08 dex.

\subsubsection{Stellar [M/H]}

Overall metallicity values, [M/H], are found within Table \ref{tab:stellarparams} and were calculated from GALAH DR3's [Fe/H] and [$\alpha$/Fe] abundances. We should note that the current TIC catalog directly incorporates GALAH DR2's iron abundances, [Fe/H], as a star's overall metallicity, [M/H]. As such, a direct comparison between such values is left to a more thorough analysis of iron abundance improvements from DR2 and DR3 in \citep{GALAHDR3}.  

\subsubsection{Stellar Radius and Mass}

From GALAH's DR3 \teff, \logg~and [M/H] measurements, alongside photometric and astrometric parameters, we have been able to constrain stellar mass, radius, age and thus luminosity values from \textsc{isochrones}. For a sanity check, we created a Hertzsprung-Russell diagram found in Figure \ref{fig:hertz} to check if our stars fell into any nonphysical areas of the diagram. A vast majority of our planet-hosting or candidate hosting stars lie on the main sequence with only 21 giant stars within our sample (Note, we classify a star as a giant, by it being cooler than 5500 K and having a surface gravity of less than 3.5; \citealt{K2HERMESsanjib}). 

Figure \ref{fig:starradmass} compares our stellar radius and mass values against other internal and external catalogs. Our stellar radius values are compared against those found within the TIC, \textit{Gaia} DR2 and the GALAH DR3 BSTEP Value Added Catalog (VAC). The BSTEP VAC results for  K2-HERMES and \textit{TESS}-HERMES stars are being released in a future paper. Briefly, stellar parameters such as age, mass and radius were computed with the BSTEP code \citep{TESSHERMES} making use of PARSEC-COLIBRI stellar isochrones \citep{PARSECiso}. BSTEP provides a Bayesian estimate of intrinsic stellar parameters from observed stellar parameters by making use of stellar isochrones. It is unsurprising, given the same data and similar methodology, that our stellar radius values agree well against those found in BSTEP, with a stellar radius bias and RMS of $-0.01^{+0.03}_{-0.04}$\rsun~and 0.04\rsun~respectively. These values are comparable to comparisons we make to the TIC and \textit{Gaia} DR2, with median biases of $-0.02^{+0.05}_{-0.08}$ \rsun~ and $-0.01^{+0.09}_{-0.07}$ \rsun~, and RMS values of 0.06 \rsun~ and 0.02 \rsun~respectively.

While GALAH DR3 does not come with stellar masses, the BSTEP VAC catalog does, such that our isochrone derived results are compared with BSTEP and the TIC values found in Figure \ref{fig:starradmass}. There are 25 stars within our sample that have no mass measurements found within the TIC (including confirmed planet hosts K2-97 b \citep{K2-97b}, NGC 2682 Sand 364 b \citep{NGCSAND364} and NGC 2682 Sand 978 b \citep{NGCSAND978}), due to the TIC only deriving masses for dwarf stars and hence no comparison between their masses are shown in Figure \ref{fig:starradmass}.

As with the stellar radii, there seems to be an overall good agreement with our derived mass values and those found within the TIC, with a bias and RMS of $0.01^{+0.11}_{-0.09}$\msun~and 0.01\msun~respectively. Our stellar mass results are comparable with stellar mass values found in GALAH's BSTEP VAC, with bias of $0.01^{+0.04}_{-0.03}$\msun~ and RMS of 0.01\msun~respectively. However, there seems to be a larger scatter for stars more massive than 1.4\msun, with our results preferentially favouring larger mass values compared to the other catalogs. These mass and radius measurements become fundamental parameters for constraining the mass and size of the planets they host, which will be discussed in the next section.

\section{Refining Planetary Systems}\label{sec:refineplanets}

With our new stellar parameters, we can now refine and redetermine the planetary or potential planetary systems that these stars host. We previously cross-matched GALAH DR3 with the NEA for known planet-hosts and K2-candidates. We also cross-matched GALAH DR3 with stars known to host TOI and CTOIs. From these cross-matches, we have 105 confirmed exoplanets, and 293 exoplanet candidates with 146 K2 candidates, 95 TOIs and 52 CTOIs. The refined planetary parameters can be found in Table \ref{tab:planetparams}.

As there are multiple catalogs with differing definitions of transit depth (i.e some report it in mmag, others as a percentage or as $(R_s/R_p)^2$), we convert all transit depths to percentages. Thus the exoplanet's radius is defined by:

\begin{equation}
    \frac{R_p}{R_\oplus} = \sqrt{0.01\Delta F}\frac{R_s}{R_\odot}\,\,,
\end{equation}

\begin{figure*}
 \centering
	\includegraphics[width=\columnwidth]{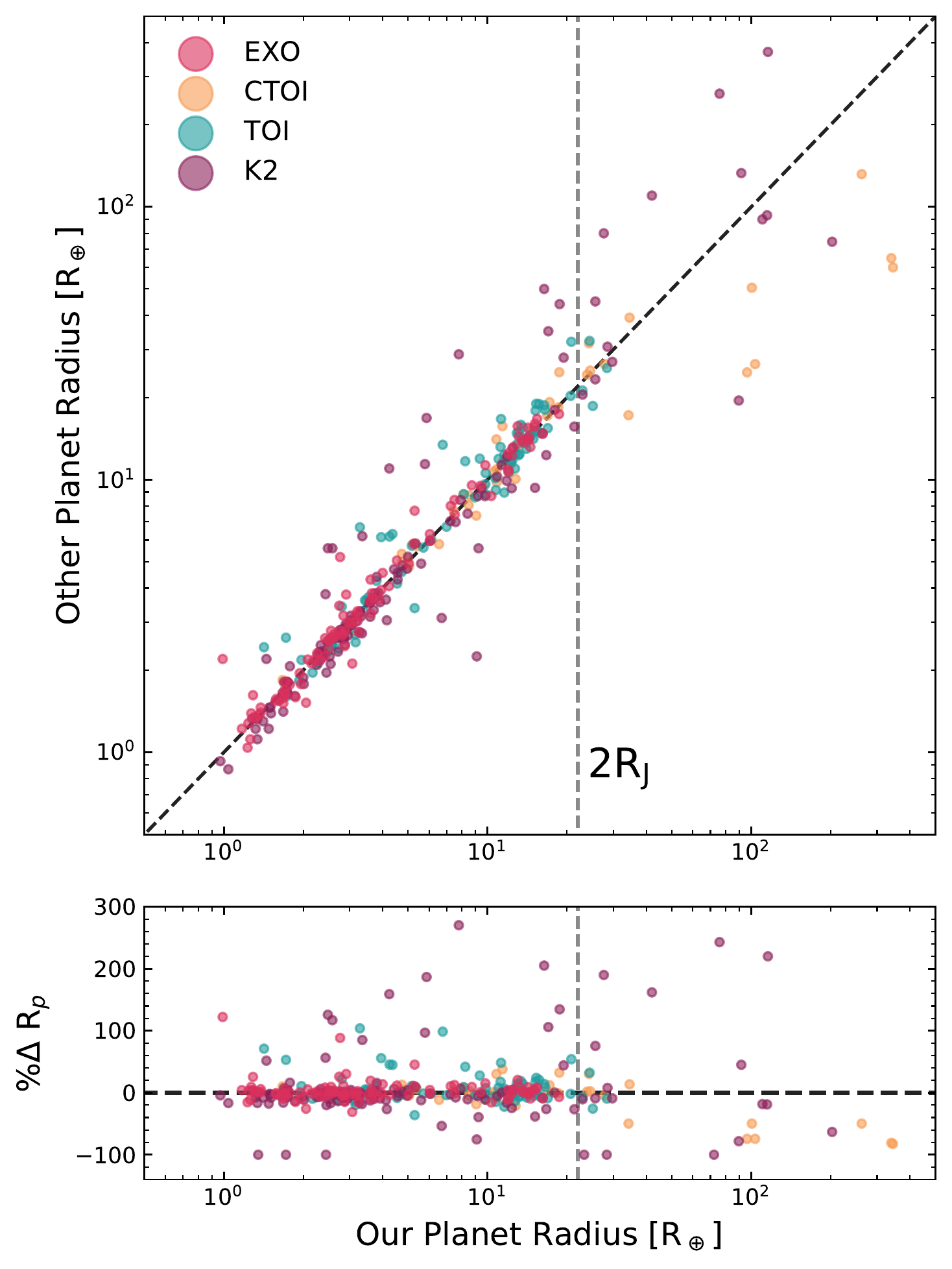}
	\includegraphics[width=\columnwidth]{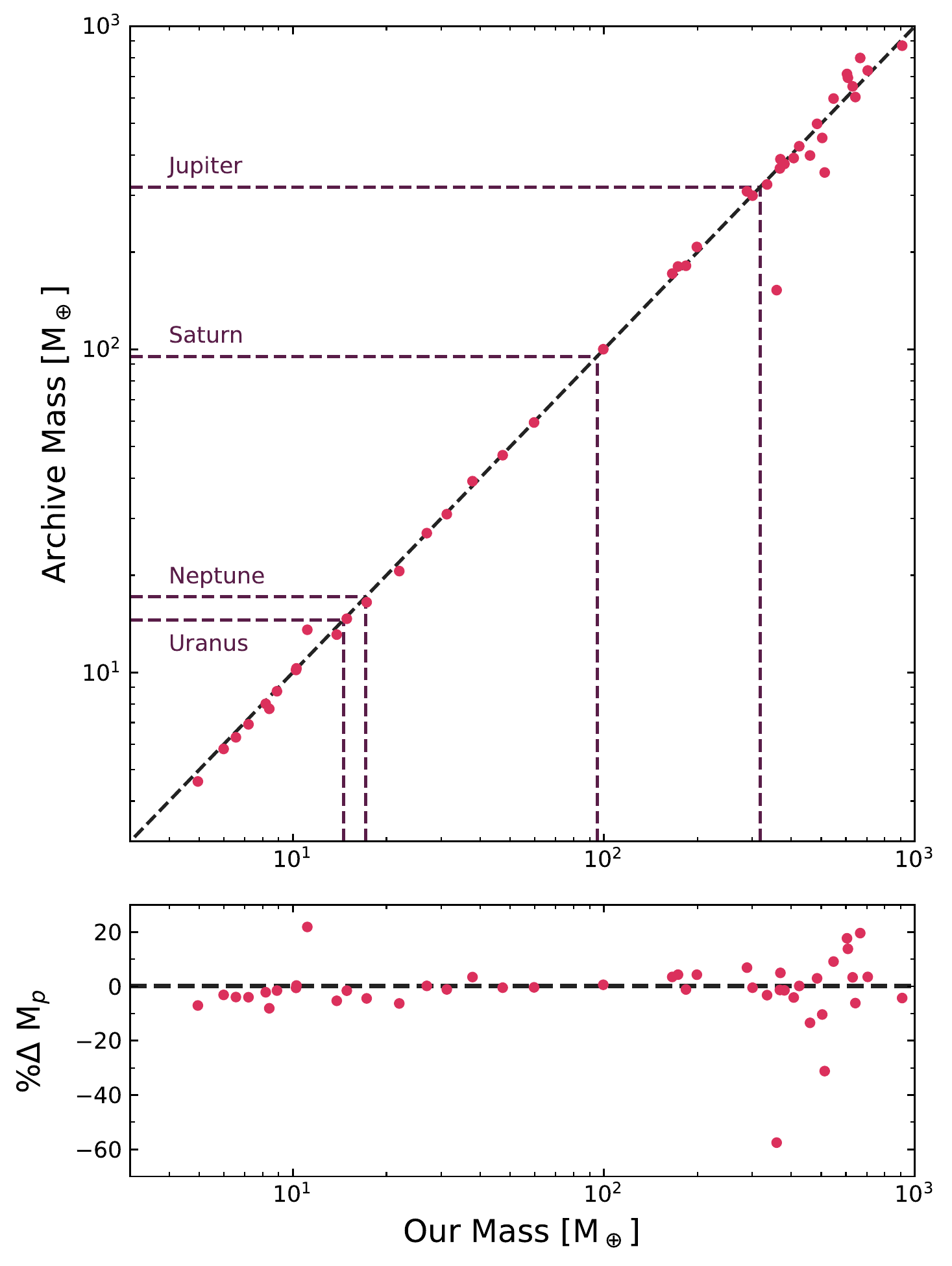}
 \caption{\textbf{Left:} Our planetary radii values are compared with literature values found within the NEA (red) K2 candidate catalog (purple) and TOI (blue) and CTOI (yellow) catalogs found on ExoFOP. No confirmed exoplanet thus far has a radius larger than twice that of Jupiter's (vertical grey-dashed line). As such, any candidate beyond the right of this line is most likely non-planetary in nature and more likely a sub-stellar or stellar companion to its host star. \textbf{Right:} We compare our planetary mass values to those found within the NEA. Lines representing the masses of Uranus, Neptune, Saturn and Jupiter are depicted as purple dashed lines across the plot. }\label{fig:planradmass}
\end{figure*}

\noindent where $R_p$ is the planet's radius, $\Delta F$ is the percentage transit depth and $R_s$ is the host star's radius. The planet's semi-major axis is calculated by Kepler's third law, incorporating our revised stellar mass values. These semi-major axis values were then used alongside stellar luminosities ($L_\star$) to derive insolation flux values, S\textsubscript{eff}, for each confirmed and candidate exoplanet. These stellar luminosities and the newly derived temperatures can then be used to determine a planet's effective temperature. If an exoplanet's atmosphere is ``well-mixed", that being there is no obvious phase function at IR wavelengths, its effective temperature will be: 

\begin{equation}
    T_p = \bigg(\frac{L_\star(1-A)}{\alpha\pi\sigma r^2}\bigg)^{\frac{1}{4}}\label{eq:equilb}
\end{equation}

\noindent where, $\alpha$ = 16, the spherical (Bond) albedo is given by A, $\sigma$ is the Stefan-Boltzmann constant and r is the star-planet separation which we've defaulted to be the semi-major axis \citep{Kane11} of the star-planet system. For a hot-dayside exoplanet, whereby its atmosphere is inefficient with respect to heat distribution, its effective temperature is calculated also using Equation \ref{eq:equilb}, but in this instance $\alpha$ = 8 instead \citep{Kane11}. Because an exoplanet's rotational period (exo-day) is currently difficult to determine, we have included both values within Table \ref{tab:planetparams}.

We have used the geometric albedos from \citet{bondAlbedoSE} with geometric albedos assumed to be 0.11$\pm$0.06, 0.05$\pm$0.04, and 0.11$\pm$0.08 for planets 1--2~\rearth, 2--4~\rearth and 4--6~\rearth , respectively. Bond albedos are then calculated from these geometric albedos using the relationship $A_B = 3A_g/2$ \citep{SeagerBook}. We also used the 1--2~\rearth~albedo value for planets $R_p < 1R_\oplus$ and used the Bond albedo value of 0.35$\pm$0.05 \citep{HJalbedo} for planets $>$6~\rearth.

For known exoplanets that have measured line-of-sight, $M_{P}\sin i$, or true mass measurements, we refine their masses given our stellar mass values given by the formula:

\begin{equation}\label{eq:rvequation}
    \frac{M_{P}\sin i}{M_{\oplus}}=\frac{K_{RV}\sqrt{1-e^{2}}}{0.0895 \mathrm{ms}^{-1}} \left(\frac{M_{*}}{M_{\odot}}\right)^{2 / 3}\left(\frac{P}{\mathrm{yr}}\right)^{1 / 3}
\end{equation}

\noindent where $K_{RV}$ is the radial velocity's semi-amplitude in ms\textsuperscript{-1}, $e$ is the orbital eccentricity and $P$ is the orbital period. This equation also assumes that M$_p$ $\ll$ M$_\star$ \citep{RVmethod}. If the orbital inclination of the system is known, we then refine the planet's true mass, $M_p$.

Because the NEA is designed now to have multiple entries for a single exoplanet or planetary system, the challenge then becomes what observables we use to refine and re-characterise these exoplanets and planetary systems. We have used a weighted mean approach to determine a single value for our K2 candidate transit depths and orbital inclinations. We have also used a weighted mean approach to derive exoplanet transit depths, radial velocity semi-amplitudes, and orbital eccentricities, inclinations and periods where available. The weighted mean of observable $X$ is calculated by:

\begin{equation}
    \bar{X} = \frac{\sum_{i=1}^{n}{w_i X_i}}{\sum_{i=1}^{n}{w_i}}
\end{equation}

\noindent where the weight of the ith data point $w_i$ is related by its error, $\sigma_i$ as $w_i = \sigma_{i}^{-2}$. The error in the standard mean is then calculated to be:

\begin{equation}
    \sigma_{\bar{X}} = \left(\sqrt{\sum_{i=1}^{n}{w_i}}\right)^{-1}
\end{equation}

\noindent These weighted mean values are then used to refine the planetary mass and radii values found in Figure \ref{fig:planradmass} with our new planetary properties found in Table \ref{tab:planetparams}.

\section{Discussion}\label{sec:compareparam}

\subsection{Identifying False Positives}

Figure \ref{fig:planradmass} shows the comparison of our planetary radii and mass values compared to catalog values found either on the NEA or NASA's ExoFOP databases. If there were multiple planetary mass or radius values, we have compared our results to the latest values within their respective databases. Using the same definition found in \citet{K2HERMESours}, we use an upper limit of R\textsubscript{P} < 2 R\textsubscript{J} $\approx$ 22\rearth~for a planetary object. We have used this cut off as there is currently no confirmed exoplanet on the archive with a planetary radius larger than this value.

There are twelve CTOIs that have radii comparable to sub-stellar and stellar objects, being larger than our 2R\textsubscript{P} limit. Of these \citet{GALAHTESS} also found the planetary radii values for CTOI 201256771.01, CTOI 201256771.02 and CTOI 300903537.01 to be too large for a planetary object. The two candidates orbiting TIC 141622065 both have radii of 344.62$\pm$6.50\rearth~and 339.57$\pm$6.40\rearth, with an orbital period separation of 0.05 days. These two candidate events are more akin to a single 3\rsun~binary star companion. Similarly, the radii for all candidates in the CTOI 91369561 system including TIC 91369561.02 and TIC 91369561.03 have radii comparable to FGK type stars. CTOI 140830390.01 and CTOI 220402290.02 have planetary radii on the cusp of the ``planetary-limit" of $\sim$22\rearth~, but was further suggested in \citet{GALAHTESS} that the candidate events within the CTOI 220402290 system are also from a single object, which is likely caused by a sub-stellar or stellar companion. CTOIs 31869740.01 and 219322317.01 are both recorded as having sub-stellar radius values of  31.61\rearth~ and 39.24$\pm$2.64\rearth~ respectively on TFOP. With our revised radius values, these two candidates also have nonphysical planetary radii, with their radii now being 24.27$\pm$2.13\rearth~ and 34.59$\pm$0.71\rearth~ for CTOI events 31869740.01 and 219322317.01 respectively. Planetary candidate CTOI 140830390.01, also known as TOI 1072.01, is more likely a brown-dwarf with a planetary radius of 27.91 $\pm$ 0.68\rearth. These results highlight the need for vetting of CTOI systems before they are made publicly available.

With our refined radii, there are five exoplanetary TOI candidates that have radii $R_p > 2R_J$. These being TOI 147.01, 565.01, 959.01, 1072.01 and 2391.01. The TESS Follow-up team contains five specialist sub-groups, with TFOP Science Group 1 (SG1) specialising in seeing limited photometry. Members attempt to confirm and refine orbital parameters from TOIs, with the project lead then up- or down-ranking the candidates' disposition depending upon their follow-up observations. Currently, TOI 147.01 and TOI 959.01 have their dispositions labelled as `False Positives', with our work also confirming that their radii alone are too big to be considered planetary in nature. TOI 565.01 was originally labelled as a `False Positive' under the TFOPWG Disposition in April 2019 but is now an `Ambiguous Planetary Candidate' as of November 2020. With an archive planetary radius of 18.62\rearth~ and now an updated radius of 25.07$\pm$0.65\rearth, this candidate is also now too large to be exoplanetary in nature and is likely an eclipsing binary event.

There are only two TOIs above our planetary radius limit, TOI 1072.01 and TOI 2391.01, which are currently labelled as Potential Candidates on TFOP, with TOI 1072.01's planetary nature already being discussed. TOI 2379.01 is on the border-line of our defined planetary radii boundary, having a radius of 21.21$\pm$1.34\rearth~ on TFOP with our radius only being slightly higher of 22.89$\pm$0.58\rearth. Notes on TFOP suggest that this event is likely an eclipsing binary, with our revised radii also being too large for a planet-like event. If confirmed to be an exoplanet, however, it would be one of the largest ever to be discovered.

There are 18 K2 candidates that have problematic radii with 13 having been found to be false positives in previous work \citep{K2HERMESours,K2HERMES2020}. Of the five K2 candidates remaining, three candidates EPIC 210769880.01, EPIC 205050711.01 and EPIC 204546592.01 all have candidate radii $>$0.35\rsun~, which far exceeds the physical radius needed to be a planetary event. There are two candidate events observed by K2 that are much closer to having radii nearer to the 2$R_J$ radius limit, these being EPIC 214611894.01 and EPIC 210598340.01.

\begin{figure}
 \centering
	\includegraphics[width=\columnwidth]{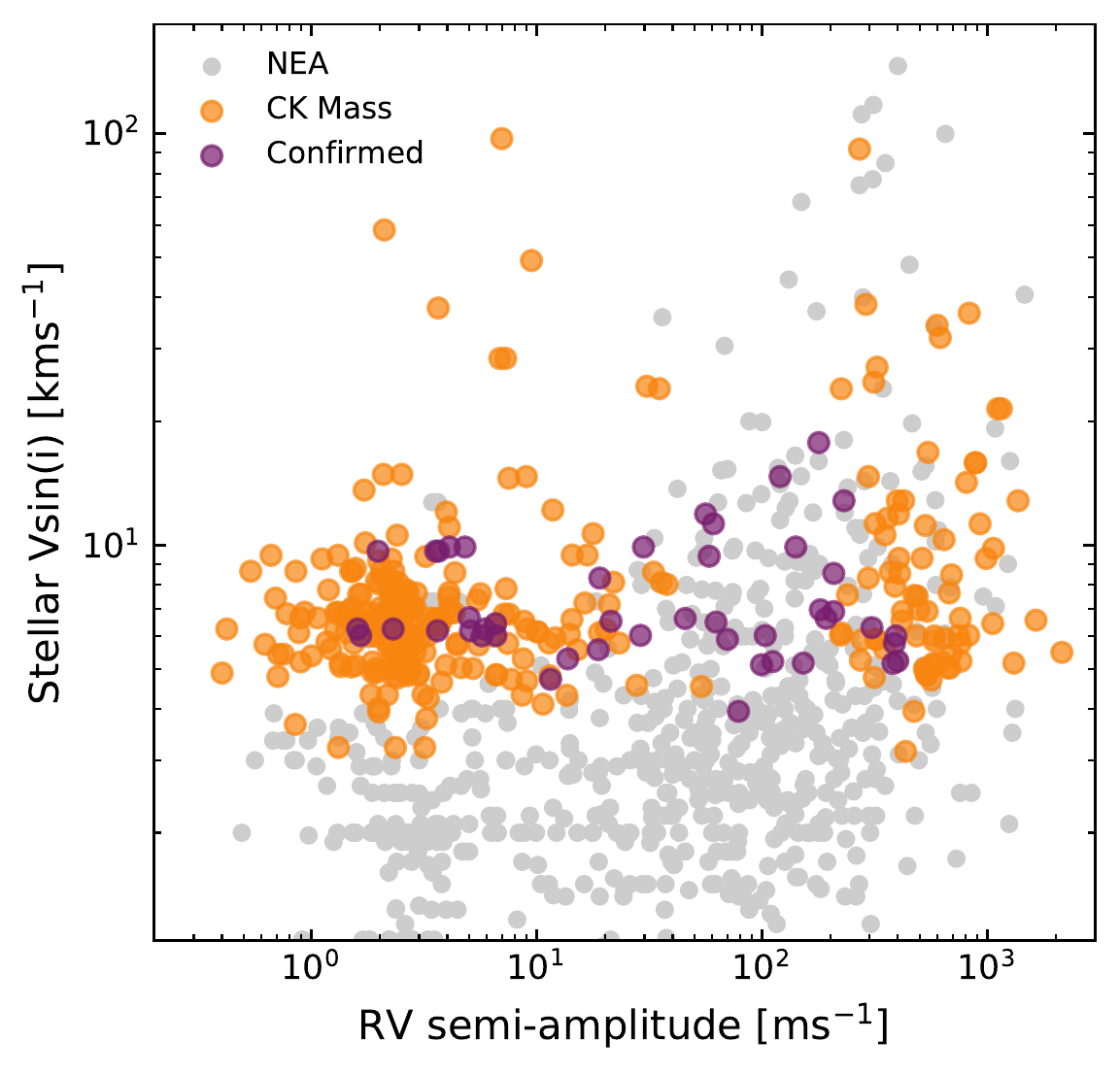}
 \caption{Expected radial velocity amplitudes for candidate and known exoplanets have been derived from expected mass values from \citet{ChenKipping} (in yellow) and compared to exoplanets with known mass values in our sample (purple) and those within NASA's Exoplanet Archive (grey)}\label{fig:vsini}
\end{figure}

EPIC 214611894.01 is a candidate event discovered through K2's seventh campaign by \citet{K2-216bA} with a candidate radius of 20.5$^{+2.8}_{-2.1}$\rearth. Its refined stellar host radius of 1.35$\pm$0.02\rsun~ is larger compared to \citet{K2-216bA}'s value of 1.21\rsun. This increase has meant the candidate's radius is now 28.50$\pm$0.63\rearth~, which is too large to be planetary in nature. A candidate first announced by \citet{EPIC2105A}, EPIC 210598340.01's radius was first determined by \citet{EPIC2105B} as being 30.71$^{+0.54}_{-6.89}$\rearth. Our newly revised candidate radius for EPIC 210598340.01 sits right on our defined radius limit, being 22.92$\pm$0.45\rearth~. This candidate radius is also smaller than the one derived by \citet{KHU20}, with their derived planetary radius for EPIC 210598340.01 of 26.764$^{1.724}_{-1.633}$\rearth. With our revised radii, it is possible that this candidate is planetary in nature, but more follow-up is needed.

\subsection{Follow up and mass confirmation from ground-based surveys}

TESS is delivering thousands of candidate exoplanets in which the community will have to confirm and characterise by both space and ground based observations. The radius measurements of these candidates will be derived mostly from TESS photometry, with confirmation mass measurements being derived from ground-based radial velocity followup. On the the biggest considerations for radial velocity followup is considering a star's rotational velocity - or in most cases, its projected rotational velocity ($v\sin{i}$). Slow rotating stars are the most preferred stellar companions, as they generally have well-defined absorption lines. As $v\sin{i}$ increases, the absorption lines needed for high-precision radial velocity measurements will broaden out, with some lines blending completely. This broadening decreases the number of well defined lines needed to obtain better radial velocity measurements, thus deteriorating the radial velocity precision. We have thus included $v\sin{i}$ values in Table \ref{tab:stellarparams} to assist followup teams in better allocating telescope time to feasible RV targets.

We have decided to forward model the likely mass and thus predicted radial velocity semi-amplitudes for confirmed and candidate exoplanets within our sample, to check which are the most viable targets for mass confirmation. To model the predicted radial velocity semi-amplitudes, we have used the \citet{ChenKipping} mass-radius relationship to derive predicted mass values for exoplanets for which there exists no current mass measurements. From this, we have used Equation \ref{eq:rvequation}, along with the host's stellar mass, the predicted planetary mass, observed orbital period and assuming circular orbits for these planets, to then obtain the predicted semi-amplitudes. These predicted semi-amplitude values are plotted against the host star's $v\sin{i}$ in Figure \ref{fig:vsini}. In Figure \ref{fig:vsini}, we have also included confirmed exoplanets with known mass measurements from NEA for comparison to our predicted RV values.

For the majority of our candidates, their host star's \vsini values are above 5\kms, however the planet's predicted RV signal is less than 5\ms. For context, only 19 confirmed exoplanets have RV signals less than 5\ms~orbiting stars with \vsini~values greater than 5\kms. We then predict that it will be difficult, with current methodology, to obtain mass confirmations for most of the smaller planet candidates orbiting stars found within our sample. In the extreme cases for smaller planets orbiting rapid-rotating stars, it will be highly improbable to derive their mass measurements.

\begin{figure*}
 \centering
	\includegraphics[width=\textwidth]{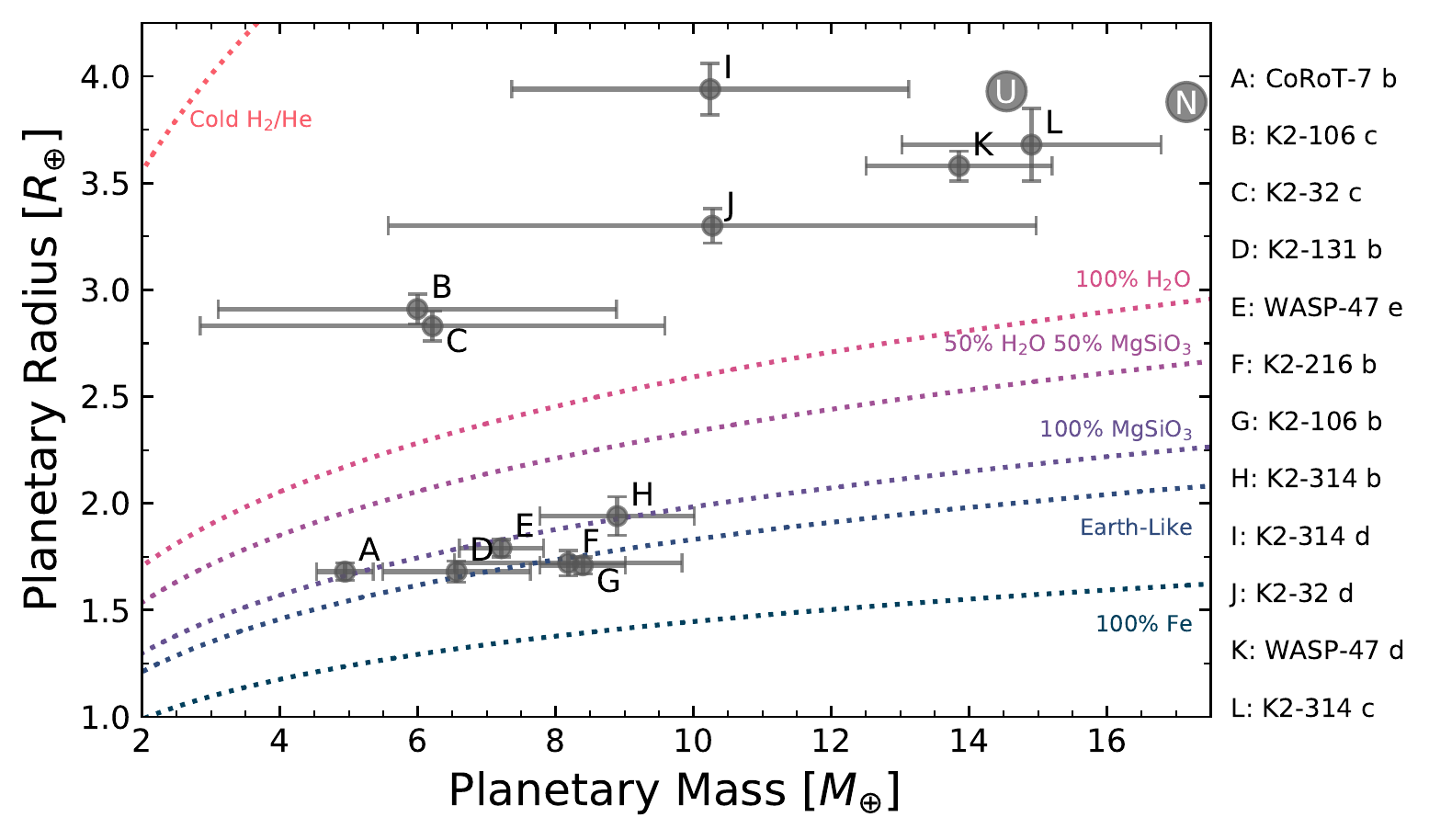}
 \caption{Revised planetary mass and radius values for confirmed super-Earth and mini-Neptune exoplanets within our sample. Each planet is labelled alphabetically, with its corresponding name to the right of the plot. 1-$\sigma$ error bars are given for each exoplanet, with Uranus and Neptune depicted on the plot as 'U' and 'N' in the plot's right-hand corner. The mass-radius relationships used for the dashed lines to show density curves for cold H$_2$/He, 100\% water, 50\% water-50\%rocky, pure rocky (containing pure post-perovskite MgSiO\textsubscript{3}), "Earth-like" (33\%Fe and 67\%rock) and pure iron worlds are from \citep{zengPREM}.}\label{fig:superearth}
\end{figure*}

For example, the star TOI-1219 (TIC 294981566,) is a rapidly rotator (\vsini~= 58.3\plusmin2.5\kms) with a 1.97\plusmin0.13\rearth~planet candidate orbiting around it every 1.91 days. We wanted to see what the expected radial velocity precision would be for such candidate orbiting around such a rapid rotator, and compare that to the expected RV signal of the exoplanet. To do this, we created a template spectrum that roughly matched the stellar properties of the host star, resampling the spectrum to match modern extreme precision radial velocity (EPRV) spectrographs (Spectral range = 400--650 nm, R $\sim$ 100,000 and 3.5 pixels per FWHM). We then recreated this stellar template for its observed \vsini~value, calculating the intrinsic error in radial velocities using \citep{Butler96}. With its current rotational velocity, the expected RV precision for TOI-1219 is 20.5 \ms, nearly an order of magnitude difference compared to the predicted RV signal of TOI-1219.01 being an estimated 2.1 \ms. Since we know what the expected RV signal of the planet candidate is predicted to be (K = 2.1 \ms) and have an intrinsic RV precision ($\sigma_{\bar{V}}$ = 20.5 \ms) we can use the formula found in \citet{Wittenmyer11}:

\begin{equation}
N = \bigg(\frac{12.3 \times \sigma_{\bar{V}}}{K+0.02}\bigg)^2\,\,,
\end{equation}

to then determine the number of observations (N) needed to detect the radial velocity signal of TOI-1219.01. In this case, the number of observations needed to confirm the mass measurement of TOI-1219.01 would be over 14,000. For context, the number of radial velocity observation we currently use for RV detection is in the order of tens or in some case nearing to hundred or two hundred observations. Thus, for a relatively hot (\teff = 6600 K) and rapidly rotating star, the precision needed to obtain a 3-sigma RV signal for TOI-1219.01 is unobtainable with current technology and methodology. Higher resolution spectrographs will be needed to determine the mass measurements of small planets orbiting rapid rotating stars. For reference, there is only one exoplanet that has a confirmed mass-measurement smaller than 4\rearth~ orbiting a star with a \vsini~greater than 20\kms, with Kepler-462 b's mass being determined through transit timing variations orbiting the rapid rotating star (\vsini = 80\plusmin3 \kms) Kepler-462 \citep{Masuda20}. Transit timing variations can therefore be another great avenue to determine the mass measurements of these candidates, rather than utilising RV measurements. Our data, along with other large stellar surveys, then provides a useful database for ground-based followup teams to determine what are the most suitable targets to follow up for mass confirmation, to maximise their resource efficiency.

\subsection{Planetary compositions from Mass-Radius Relations}

\subsubsection{Our results}

\begin{figure*}
 \centering
	\includegraphics[width=\textwidth]{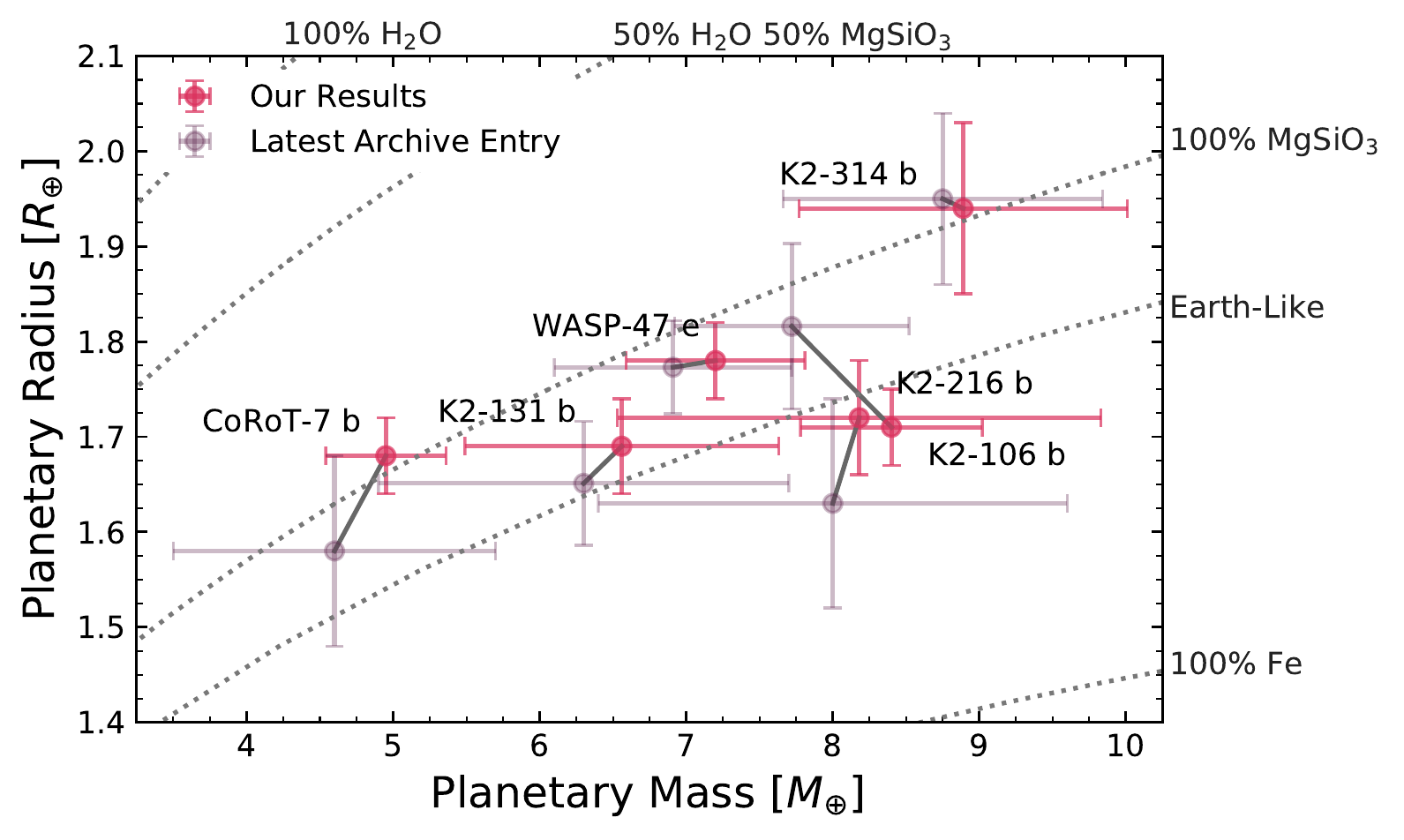}
 \caption{Combining the weighted mean and updated stellar parameters approach of our work (labelled as red crosses) has improved the mass and radius errors on several exoplanets that are thought to be rocky in nature within our work. The latest entry within the NEA is shown for reference as grey crosses.}\label{fig:superearthzoom}
\end{figure*}

Within GALAH's latest data release, there are 105 confirmed planet-hosting stars. There was only one planet that was omitted in our analysis, this being the exoplanet Pr0211 c, as its orbital period error was too large to facilitate in the MCMC analysis of our work. With our previous line of thought, there are no planets with revised radii measurements that exceed the 2$R_J$ limit. Nor do any revised mass measurements, where available, exceed the theoretical exoplanet mass limit of $\sim$13$M_J$ \citep{planetmasslimit}. Figure \ref{fig:planradmass} shows our newly derived mass values compared to those found in the literature. We should note that in Figure \ref{fig:planradmass}, the comparison of literature values are median mass and radius values taken from the NEA. There are three visible outliers within the comparison mass plot, CoRoT-7 c (11.12$\pm$0.79\mearth), K2-97 b (359.26$\pm$38.56\mearth) and K2-34 b (512.48$\pm$17.39\mearth). With further inspection, these three outliers are mainly caused by various anomalies within the NEA, rather than drastic changes with the refined stellar parameters, for example, K2-97 b.

An example of the caveats of both our methodology and utilising any form of heterogeneous archival data comes from the revision of planetary system K2-97. First discovered by \citet{K297bA}, K2-97 b is a Jupiter-sized exoplanet that had an initial radial velocity semi-amplitude of 103$\pm$8~ms$^{-1}$, inferring a planetary mass of 350$\pm$35\mearth, where our refined value is 359.26$\pm$38.56\mearth.

However, its radial velocity semi-amplitude was later revised by \citet{K297bB} through additional radial velocity measurements to be 42.1$^{+4.3}_{-4.2}$~ms$^{-1}$, a decrease of over 56\%. These new measurements imply a revised mass of 153$\pm$22\mearth. But this RV value is not included within the NEA, and hence wasn't used within our methodology until now.

If these literature values were treated as independent measurements, our weighted mean approach would yield an estimated mass for K2-97 b of 146.7$\pm$18.4\mearth. But, if you then incorporate both of these values in our weighted mean approach, the mass of K2-97 b then becomes 194.54$\pm$19.54\mearth. Additionally, if we were only to use the default parameters (\textsc{default\_flag}) used by the NEA, K2-97 b's parameters would include its planet radius but no information would be present in regards to its planetary mass \citep{K2-97b}.

With 469 independent records for our sample of 105 exoplanets within the NEA, it would be inefficient to independently review every record entry on the archive. The community can only utilise the information that is present within its current iteration. As the field becomes more reliant on heterogeneous data to infer exoplanet populations and demographics, it is crucial that all of these data are accounted for. Hence, there needs to be a discussion within our community on making data and fundamental parameters within our literature more accessible and refine our best practice on updating information on exoplanetary systems.

Because mass and radius measurements of an exoplanet are derived through independent techniques, there is a small overlap for exoplanets with both mass and radius measurements within our sample. We can plot some of our 39 exoplanets with both inferred mass and radius values, onto a mass-radius diagram. In Figure \ref{fig:superearth}, we have plotted all of the exoplanets within our sample with a planetary radius, R\textsubscript{p} $\leq$ 4\rearth, with known mass and radius measurements. Even within our relatively small sample, there is a large variation of planetary mass for a particular planetary radius, showing the compositional diversity of known exoplanets. There are two distinct exoplanet groups contained with this phase space, sub-Neptunes (i.e WASP-47 d, K2-314 c etc.) and super-Earths (i.e CoRoT-7 b, K2-106 b etc.).

Of these 12 exoplanets, six have radii which put them within the exoplanet category known as super-Earths. All of these super-Earths either straddle or sit within the super-Earth radius gap, a scarcity of planets with radii, 1.5\rearth~$\le$ R\textunderscore{p} $\le$ 2\rearth~\citep{FultonLumps}, indicating that these planets may be transitioning between their radii being dominated by their rocky bodies, rather than gaseous envelopes. We show in Figure \ref{fig:superearthzoom} a zoom-in of our refined mass and radius values, compared to the latest entry in the NEA. There are some cases, where there are only single values for a given exoplanet's mass and radius value (e.g K2-314 b) where there is a shift in mass and radius values given our change in stellar values. But there are also several cases here where we have combined multiple mass and radius measurements within the archive, as well as our refined stellar values, to better refine the characteristics of these known super-Earths (e.g CoRoT-7 b and K2-106 b). More detail for each of these planets is discussed below.

\subsubsection{Fundamental Properties of our Ultra-Short Period Exoplanet Sample}

(i) \textbf{K2-314 b:} K2-314 b, also known as EPIC 249893012 b, is a super-Earth sized exoplanet first discovered in 2020 by \citet{K2piSystem} orbiting a slightly evolved, metal-rich ([Fe/H] = 0.19) G-type star. Our stellar parameters for K2-314, 1.71\plusmin0.03\rsun and 1.07\plusmin0.03\msun, compared to those found in \citet{K2piSystem}, 1.71\plusmin0.04\rsun and 1.05\plusmin0.05\msun, are almost identical with only a very slight adjustment to the star's mass values. This small discrepancy in stellar values translates to an insignificant change in the exoplanet's mass and radius values, with our values of 1.94\plusmin0.09\rearth and 8.89\plusmin1.12\mearth for K2-314 b, compared to its original values of 1.95$^{+0.09}_{-0.08}$\rearth and 8.75$^{+1.09}_{-1.08}$\mearth \citep{K2piSystem}. This small discrepancy is also found with its exoplanetary siblings K2-314 c and K2-314 d, found in Figure \ref{fig:superearth}.

(ii) \textbf{K2-106 b:} Found within the Pisces constellation is a two-planet system, orbiting the GV star K2-106, also known as EPIC 220674823 \citep{K2-106}. One of these planets is the ultra-short period (P = 0.567 days) super-Earth K2-106 b. Combining multiple radial-velocity and photometric measurements, our revised planetary radius and mass values for K2-106 b are 1.71\plusmin0.04\rearth~and 8.39\plusmin0.62\mearth~respectively. These revised planetary radii and mass values now place K2-106 b underneath the density curve of a planetary body with a density like that of Earth.

By far, the most interesting aspect of K2-106 b is the amount of incoming insolation flux being received on its surface, 4330 times that of Earth. K2-106 b's well mixed and hot day-side equilibrium temperatures of 2160~K and 2570~K respectively, far exceeded the condensation temperatures of most refractory elements. The day-side equilibrium temperature would be more akin to that on the surface, as tidal locking would become likely with an orbit so short. With an equilibrium temperature of 2568\plusmin74~K, our newly revised parameters for K2-106 b makes it the hottest known super-Earth to date. There are only eight exoplanets on the NEA that have higher equilibrium temperatures. However, as shown by \citet{SupEarthEscape}, ultra-hot worlds like K2-106 b, WASP-47 e and CoRoT-7 b (discussed later in this section) could retain atmospheres, even atmospheric constituents such as O, H\textsubscript{2}O and CH\textsubscript{4}, given their relatively high escape velocities. K2-106 b's escape velocity is 24.76\plusmin0.97~kms\textsuperscript{-1}, comparable to Neptune's escape velocity of 23.5~kms\textsuperscript{-1}. 

(iii) \textbf{K2-216 b:} Also contained within the Pisces constellation is the K5 V dwarf, K2-216, with a single super-Earth companion, K2-216 b \citep{K2-216bA,K2-216bB}. Our stellar radius and mass values for K2-216 are consistent with other surveys, them being 0.69\plusmin0.01\rsun~and 0.72\plusmin0.02\msun~respectively. As with K2-106 b, the planet is also on an ultra-short orbit of nine hours, with a single radial velocity \citep{K2-216bC} and several transit detections used within our methodology to revise its fundamental parameters. We have revised the planetary radius and mass of K2-216 b to be 1.72\plusmin0.06\rearth~and 8.18\plusmin1.65\mearth~respectively. With K2-216 b's relatively high density (8.85\plusmin1.99 gcm\textsuperscript{-3}), high escape velocity (24.38\plusmin2.55~kms\textsuperscript{-1}) and moderate equilibrium temperature (1217\plusmin34~K), it can be capable of an atmosphere comprised of H\textsubscript{2}O, N\textsubscript{2} and CO\textsubscript{2} \citep{SupEarthEscape}.

(iv) \textbf{WASP-47 e:} Contained within a four-planet system is a super-Earth exoplanet known as WASP-47 e, first discovered in 2015 \citep{WASP47eA,WASP47eB}. Orbiting around a iron-rich ([Fe/H] = 0.45\plusmin0.09 dex) G9V dwarf every 0.78 days, this exoplanet, and its companions have been characterised by numerous studies since its discovery \citep{WeissWasp,DornCaAlPaper,Wasp47Dark}. With our refined stellar parameters (1.13\plusmin0.02\rsun, 1.06\plusmin0.05\msun) and combining multiple independent archive observations, we have determined the radius and mass of WASP-47 e to be 1.79\plusmin0.04\rearth~and 7.21\plusmin0.61\mearth~respectively.

\begin{figure*}
	\includegraphics[width=\columnwidth]{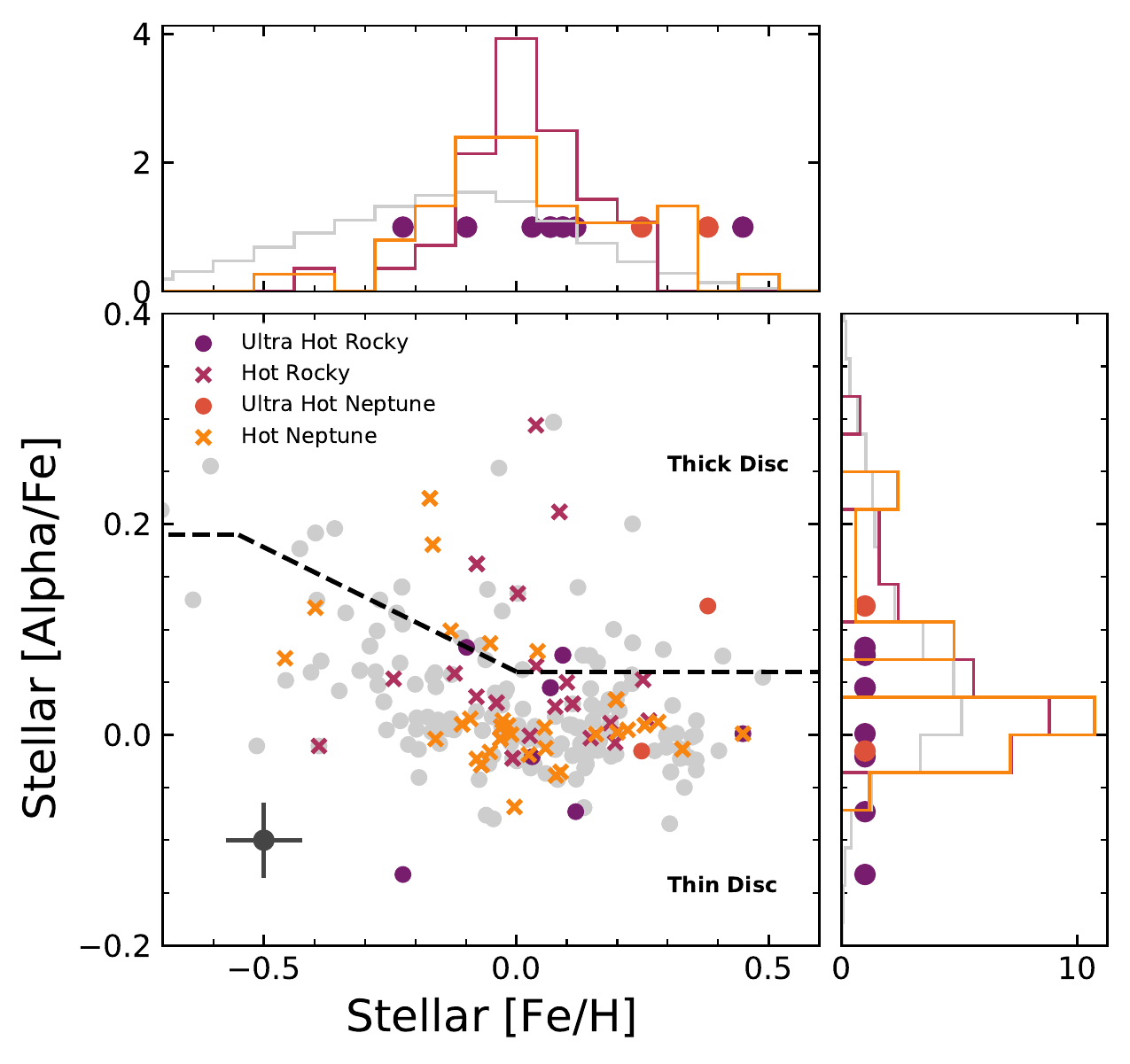}
	\includegraphics[width=\columnwidth]{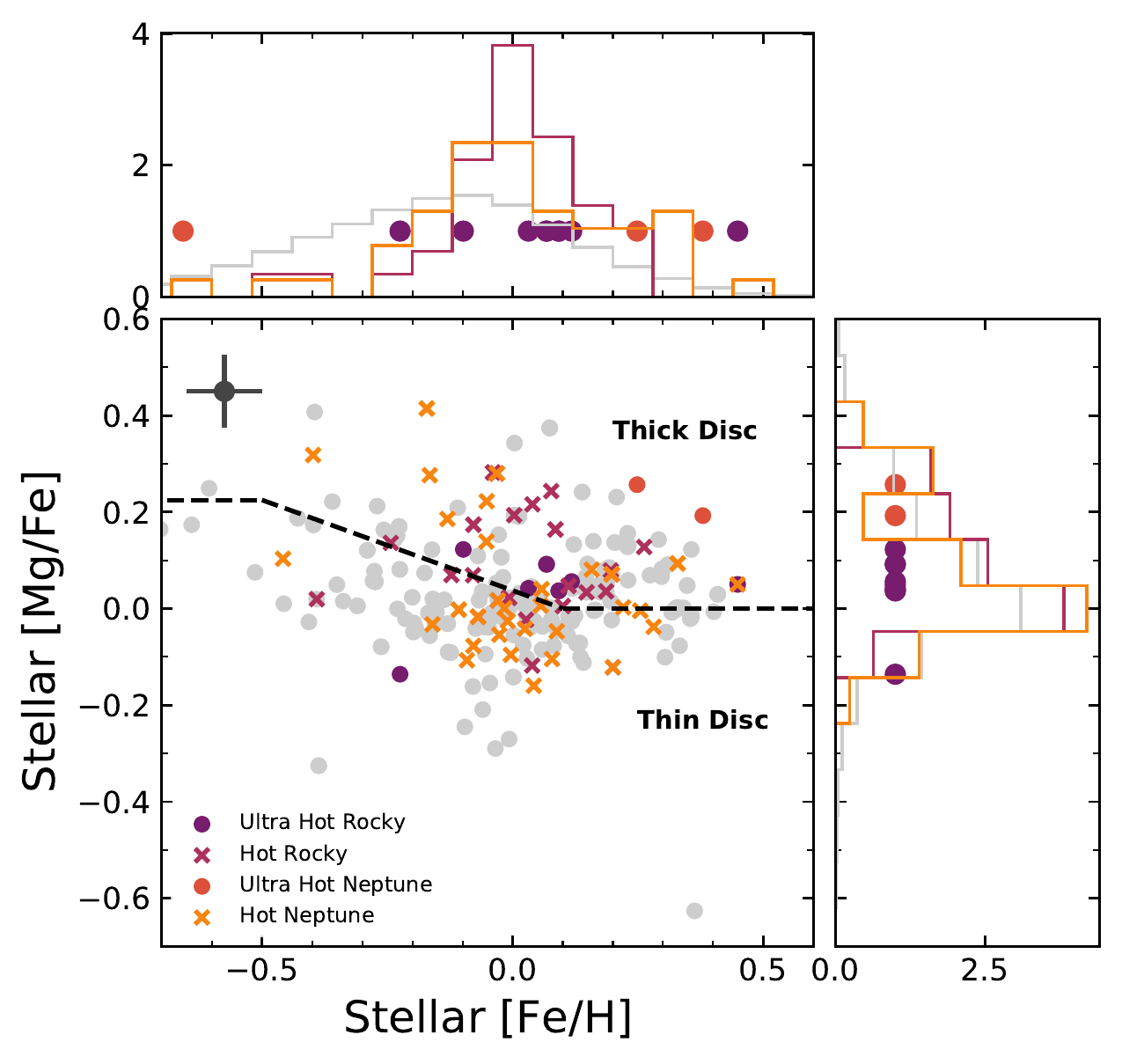}
	
 \caption{Left: Stellar iron abundance values are plotted against their $\alpha$ (left) and magnesium (right) abundances for our confirmed and candidate planet host stars. The dashed-lines represents the distinction between the different stellar populations of thick and thin-disc stars as shown in \citet{Adibekyan12HARPS} and \citet{Hayden2017} for our $\alpha$ and Mg abundances respectively. Different short (terrestrial - pink cross; Neptune - yellow cross) and ultra short (terrestrial - purple disc; Neptune - orange disc) period planet populations are highlighted within the figure. Thick disc stars are typically associated with being above the plane of the dark grey dashed line, with thin-disc stars being associated below it. Median errorbars are shown within each plot as grey markers.}\label{fig:alpha}
\end{figure*}

Its equilibrium temperature rivals that of K2-106 b, with our derived T\textsubscript{eq} being 2514\plusmin70~K. \citet{DornCaAlPaper} has postulated that WASP-47 e may have an exotic composition of a Ca and Al-rich interior without any atmosphere. But if there is an atmosphere, with an escape velocity of 22.44\plusmin0.99 \kms, it may contain ionic calcium, silicon, magnesium, and maybe even aluminium with its extreme equilibrium temperature \citep{DornCaAlPaper}. The albedo of such worlds is still a field of active research, with \citet{Wasp47Dark} being able to constrain the geometric albedo of its sibling, WASP-47 b, to be 0.0016 but were unable to constrain WASP-47 e's. However there was some evidence to suggest that WASP-47 e's geometric albedo could also be as low as WASP-47 b, but a wide range of geometric albedo solutions were also found. Contrary to this, \citet{AlbedoSupEarth} found that high spherical albedos for super-Earths could be explained by molten Fe-rich magma oceans on the surface of ultra-short period worlds \citep{AlbedoSupEarthExamp}. 

(v) \textbf{K2-131 b:} Orbiting around the K dwarf K2-131 is an ultra-short period exoplanet known as K2-131 b \citep{k2-131b}. Currently a single planet system, our stellar mass and radius values of 0.75\plusmin0.01\rsun~and 0.82\plusmin0.02\msun~are comparable to values found within the archive. Discovered in 2017, we have combined our stellar parameters and observed parameters from other surveys to revise K2-131b's radius and mass to be 1.68\plusmin0.05\rearth~and 6.56\plusmin1.07\mearth~respectively. K2-131 b is another super-Earth predicted to host an ocean of molten lava, with a hot-dayside equilibrium temperature of 2496\plusmin66 K. Its density, escape velocity and insolation flux of 7.52\plusmin1.37~gcm\textsuperscript{-3}~, 22.09\plusmin1.86~kms\textsuperscript{-1}~and 3865\plusmin300~S\textsubscript{$\oplus$} respectively, is comparable to that of K2-141 b \citep{K2-141}. Theoretical models have predicted that K2-141 b could have an atmosphere of Na, Si and SiO\textsubscript{2} that is continually replenished from the up-cycling of vapourised refractory material being displaced back into its atmosphere \citep{supEarthAtmos}. Such atmospheres could potentially be detected through space-based telescopes such as JWST, which will be launched later this year.

(vi) \textbf{CoRoT-7 b:} Whilst there have been smaller exoplanets discovered since, CoRoT-7 b \citep{corot7bA,corot7bB} was one of the first ultra-short period super-Earths to be discovered, the origins and evolution of which are still an active area of research \citep{Leger2011,Winn2018,Dai2019,Lichtenberg2021}. We have determined the stellar radius and mass of CoRoT-7 to be 0.84\plusmin0.01\rsun~and 0.87\plusmin0.03\msun~respectively. Given the observed transit depths and semi-amplitudes of CoRoT-7 b found within the literature, and our derived astrophysical parameters for CoRoT-7, we have refined its planetary radius and mass to now be 1.68\plusmin0.04\rearth~and 4.94\plusmin0.41\mearth. That is, we have increased its radius and mass precision to 2.4\% and 8.3\% respectively, the most precise physical values for CoRoT-7 b currently available.

With these precise values, CoRoT-7 b's radius is now large enough for an atmosphere to contribute to its overall radius \citep{Rogers2015_16earth}, with its density being, 5.73\plusmin0.61 gcm\textsuperscript{-3}, comparable to that of Earth's, $\rho$\textsubscript{$\oplus$} = 5.51~gcm\textsuperscript{-3} \citep{ExoplanetHandbook}. Receiving 1682\plusmin122 S\textsubscript{$\oplus$}, CoRoT-7 b's well-mixed and hot day-side equilibrium temperatures of 1705 K and 2027 K make it one of the cooler ultra-short period planets discussed in this section.

\subsection{Chemical Abundances of Confirmed and Candidate Exoplanets}

One of the biggest strengths of all-sky surveys such as GALAH is the not only access to the physical properties of these stellar hosts within our sample, but also the numerous chemical abundances that are included as well. The chemical links between stars and the planets that they host has been widely studied \citep{AdibekyanHeavyMetal,hotjupfeh05,Teske_metalrichhosts}. We are interested within our sample if there are any thick-disc hosts within our sample, and to see if there is a trend between stars that hosts close in gaseous worlds, compared to their rocky hosting counterparts.  

\subsubsection{Searching for thick-disc hosts within GALAH}\label{sec:thicc}

The density and populations of stars vary significantly within the Milky Way's disc. The thin-disc is contained within the galaxy's innermost plane, hosting relatively young (6 Gyr), iron-rich ([Fe/H]\textsubscript{thin} $\sim$ 0.0 dex), $\alpha$-poor ([$\alpha$/Fe]\textsubscript{thin} $\sim$ -0.1 dex) and low total space velocity ($v_{tot} \leq$ 50~kms\textsuperscript{-1}) stars \citep{K2HERMESsanjib,Nissen2004}. However the thick-disc, lying in the Milky Way's outer plane, consists of a much higher proportion of older stars, with a mean stellar age around 9.5 Gyr \citep{K2HERMESsanjib}. Not only are these stars older, but their iron abundance is lower ([Fe/H]\textsubscript{thick} $\sim$ - 0.367 dex ), their $\alpha$-process (i.e. Mg, Si, Ca and Ti) elemental abundances are enriched ([$\alpha$/Fe]\textsubscript{thick} $\sim$ 0.218 dex), and have faster velocities (70 $\leq v_{tot} \leq$ 200~kms\textsuperscript{-1}) compared to their thin-disc counterparts \citep{K2HERMESsanjib,Nissen2004}. Within the Solar neighbourhood, 1\% to 12\% of stars are estimated to be considered members of the Milky Way's thick-disc \citep{Bland-Hawthorn16}, with a few exoplanets being announced orbiting thick-disc stars \citep{thiccLHS1815b,thiccTOI561,thiccWASP21b}. With the above population statistics, there should be in the order of 35 -- 400 exoplanet hosts from the thick-disc in which we can then better understand the underlying exoplanetary population and architectures of thick-disc planetary systems.


With this motivation, we used GALAH DR3 stellar abundances, specifically the iron and $\alpha$ abundances, along with the galactic kinematic and dynamic information within GALAH's value added catalog, to constrain the stellar populations of known and candidate exoplanetary systems. In Figure \ref{fig:alpha} we plot the iron abundance of known confirmed and candidate systems against the stars' $\alpha$ and magnesium abundances (a more thorough discussion between the populations outlined in the Figure is discussed in Section \ref{sec:popdiff}). Figure \ref{fig:alpha}A shows a host star's iron abundance against its $\alpha$ abundance, with dashed lines within the plot chemically separating stars within the thick and thin disc regions of the Milky Way as per \citet{Adibekyan12HARPS}. We also use a one-sigma cutoff of the $\alpha$ and iron abundances from the separation line to determine the stellar populations from these chemical abundances. From these cut offs, we have 13 stars, or 5\% of our host stars potentially from the thick-disc. These stars, along with their $\alpha$ and iron abundances can be found within Table \ref{tab:thicc}. 

\begin{figure*}
	\includegraphics[width=\columnwidth]{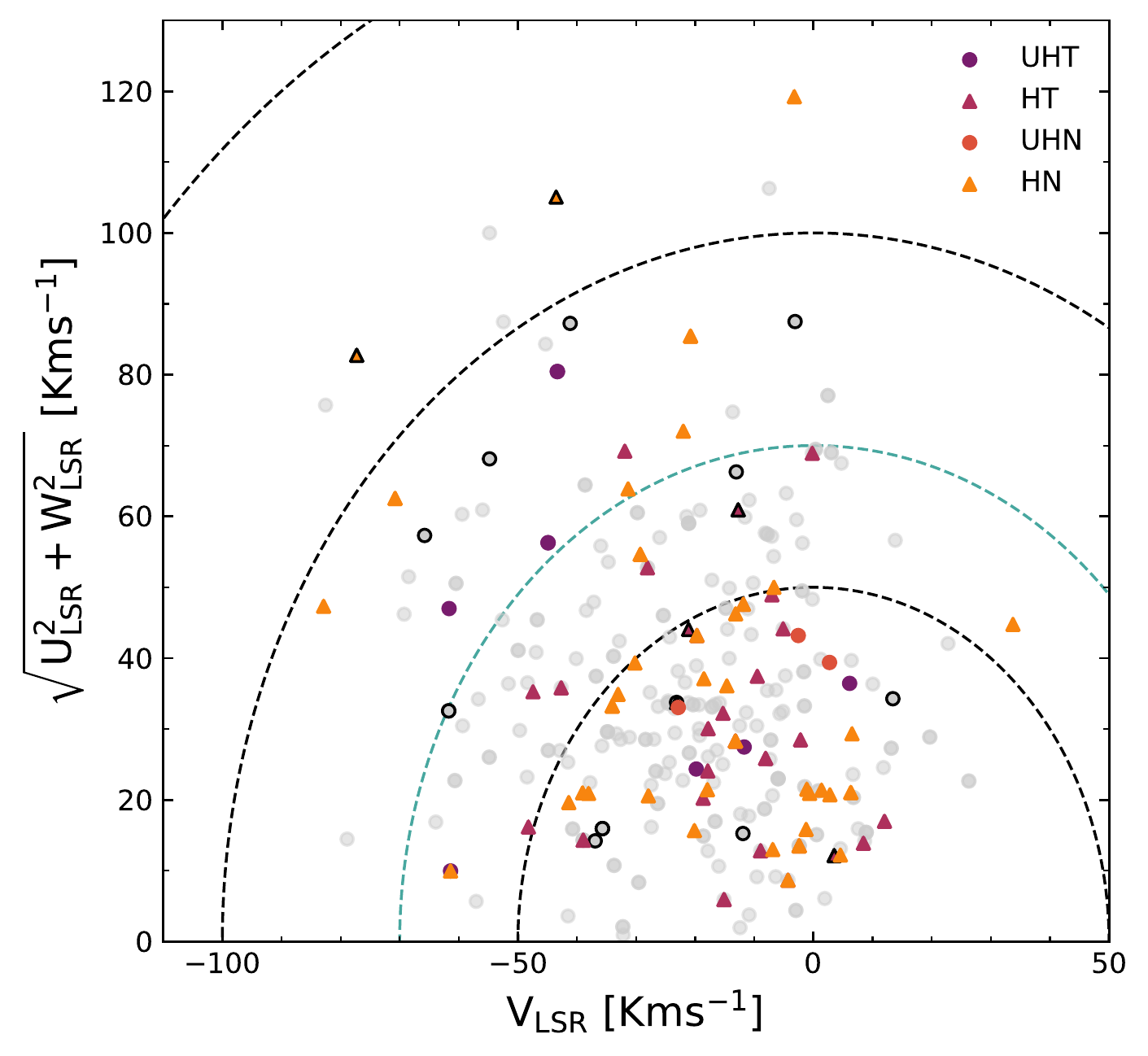}
	\includegraphics[width=\columnwidth]{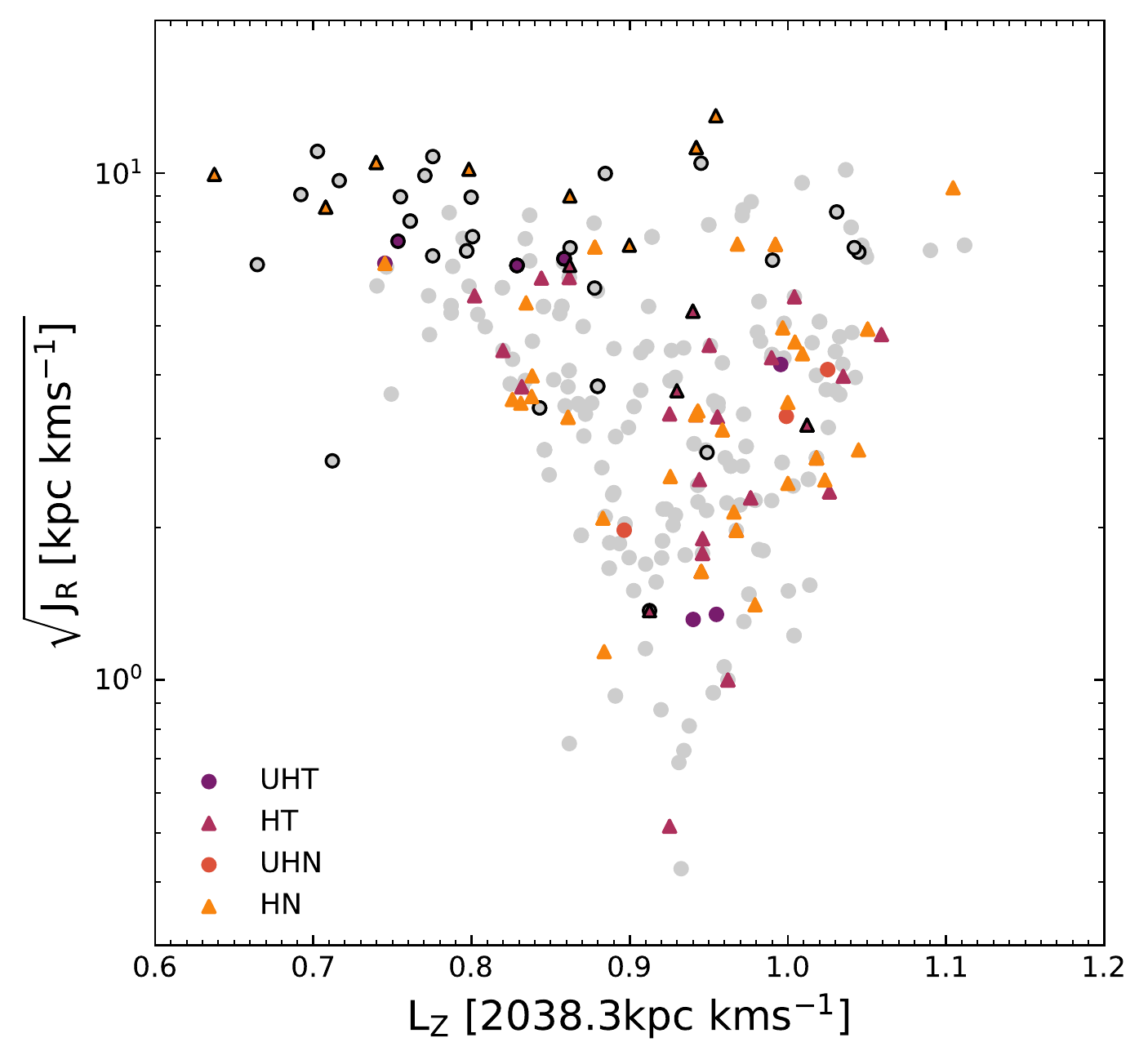}
	\caption{Left: Toomre diagram for the planet hosting and stars without planets. The red squares and blue triangles represent stars with Jupiter- and Neptune-mass planets, respectively. The magenta circles and green asterisks refer to the high-$\alpha$ metal-poor (chemically defined thick disc) and high-$\alpha$ metal-rich stars without planets, and the black dots refer to the chemically defined thin disk non-host stars. Dotted grey lines indicate total space velocities of 50, 100 and 150\kms~with the teal dashed-line of 70\kms~being our thick-thin disc kinematic cutoff line. Stars with total velocities greater than 70\kms~ are also outlined in the right figure. Right: Stellar dynamics of our planet-hosting stars including the dynanic actions J\textsubscript{R} and L\textsubscript{z}. L\textsubscript{z} of the Sun is determined to be 2038.3 kpc \kms \citep{GALAHDR3}. \label{fig:toomre}}
\end{figure*}

Another way to determine the stellar populations of stars within the galaxy is through kinematic and dynamic data. A stars velocity towards the Galactic Centre (U), in the direction of rotation (V) and upwards from the disk (W) can be determined though large spectroscopic surveys combining astrometric data from \textit{Gaia} \cite{GAIAEDR3} and radial velocity data from GALAH \citep{GALAHDR3,Bland-Hawthorn16}. We have used these stellar velocities from GALAH DR3 to create a Toomre diagram found in Figure \ref{fig:toomre}. The Toomre diagram shows the the radial and perpendicular ($\sqrt{U_{LSR}^2 + W_{LSR}^2}$) stellar velocities for all known confirmed and candidate exoplanet hosts, corrected for the Local Standard of Rest (LSR). We have included in Table \ref{tab:thicc} stars will total space velocities ($v_{tot} = \sqrt{U_{LSR}^2 + V_{LSR}^2 + W_{LSR}^2}$) greater than 70 \kms to bring the number of potential thick-disc host stars to 30. We can also see in the Figure, that there are stars that have chemical abundances suggesting that they are from the thick-disc, yet have stellar velocities similar to that of thin disk stars. However, past surveys have also shown that galactic thick-disc stars will often have similar kinematics to those stars found in the thin-disc \citep{PAST,Kovalev19}.

We have also included in Figure \ref{fig:toomre}, the stellar dynamics of our exoplanet hosts; L\textsubscript{z} and J\textsubscript{R}. L\textsubscript{z} is the azimuthal angular momentum of a star and describes the amount of rotation a star's orbit has around the Galactic centre \citep{Trick19,Bland-Hawthorn16}. J\textsubscript{R} is describes the radial action of a star and can be considered a measure of a star's orbital eccentricity \citep{Trick19,Bland-Hawthorn16}. Thick-disc stars would be considered to have higher J\textsubscript{R}, as thick-disc stars have more eccentric orbits than their thin-disc counter parts \citep{GALAHDR3} and have L\textsubscript{z} values diverging away from the solar neighbourhood L\textsubscript{z} value of 2038.3 kpc \kms.

\citet{Carrillo20} collated a large chemo-kinematic database for stars being observed by \textit{TESS}, and determined probable likelihoods of stars within their database to be members of the thick-disc (TD/D) as per \citet{Bensby14}. These probabilities determined by \citet{Carrillo20} suggest that stars with a TD/D > 2 would be likely members of the thick-disc, with stars having TD/D < 0.5 being associated more with the thin-disc and stars in between these values being transitional-disc stars. We have cross-matched \citet{Carrillo20}'s catalog, with our thirty thick-disc candidate stars, with 18 confirmed cross-matches. All of the above information can now better inform us of what stellar populations these planet hosts are likely to be a part of.

From the above information we have four stars including confirmed planet hosts NGTS-4, K2-183 and EPIC 211770696 along with the multi-planet hosting candidate star TOI-810 to be members of the thick-disc. \citet{NGTS4b} announced the discovery of NGTS-4 b, showing that its host star's kinematics made it a member of the thick-disc. With GALAH's chemical abundances, [Fe/H] = -0.17 \plusmin 0.06, [$\alpha$/Fe] = 0.22 \plusmin 0.03, we independently confirm that NGTS-4 is indeed a member of the thick-disc. With the highest J\textsubscript{R} within our candidates, along with a low iron abundance ([Fe/H] = -0.44 \plusmin 0.18) and enriched $\alpha$ abundance ([$\alpha$/Fe] = 0.07 \plusmin 0.1) we also consider candidate planet host EPIC 211064647 to also be a member of the thick disc. It is interesting to note that TOI-810 has one of the lowest J\textsubscript{R} values within the potential thick-disc members, however has a TD/D value of 2.53. This shows that all chemo-kinematic and dynamical information needs to be considered before determining what population a star is associated with. With \citet{Carrillo20}'s database and our results, we also consider EPIC 211800191 and EPIC 213546283 to be members of the transition area of the Milky Way. During the writing of this paper, EPIC 211770696.01 and EPIC 211978988.01 were confirmed as exoplanets and were given the designation of K2-337 b and K2-341 b respectively \citep{deLeon21}. However, our work shows for the first time that K2-337 b has characteristics associated with being a thick-disc planetary system and K2-341 b is orbiting a star likely associated with the Milky Way's thin disc. Due to the complex and non-linear nature that the chemical, kinematic and dynamic information that stars have with their associated stellar groups, we leave the rest to be characterised in further studies. Having a homogeneous and more inclusive sample of thick-disc stars, such as \citet{PAST} and others, would allow exoplanetary scientists to better probe the characteristic planetary population differences between the stellar populations, leading onto implications for astrobiology and habitability across the Milky Way \citep{Santos17}.

\begin{table*}
\caption{Our potential thick-disc host stars are presented in this table with their chemical, kinematic and dynamic information. The Table is sorted by the star's radial action. TD/D probability values are taken from \citet{Carrillo20} for stars within our sample that had matching TIC IDs.}
\begin{center}\label{tab:thicc}
\begin{tabular}{lccccccc}
\hline
Star name & [Fe/H] & [$\alpha$/Fe] & V\textsubscript{SLR} & $\sqrt{U_{SLR}^2 + W_{SLR}^2}$ & J\textsubscript{R} & L\textsubscript{Z} & TD/D\\
& dex & dex & \kms& \kms& kpc~\kms & kpc~\kms & \\
\hline
EPIC 211064647& -0.44 \plusmin 0.18 & 0.07 \plusmin 0.1 & -259.1 & 79.94 & 715.6 & -91 \\
K2-64 & -0.08 \plusmin 0.12 & 0.09 \plusmin 0.07 &-3.241&119.2&168.4& 1935 \\
K2-181 & 0.25 \plusmin 0.07 & 0.01 \plusmin 0.04 &-20.8&85.45&126.2& 1910 & 0.221 \\
EPIC 211770696& -0.36 \plusmin 0.09 & 0.2 \plusmin 0.05 &-82.53&75.71&121.8& 1425 & 4.52 \\
TOI-933 & -0.61 \plusmin 0.06 & 0.26 \plusmin 0.03 &-54.8&68.13&116.1& 1573 & 0.291 \\
K2-204 & -0.11 \plusmin 0.06 & 0.01 \plusmin 0.02 &-70.77&62.55&110.2& 1500 \\
EPIC 212624936& -0.07 \plusmin 0.08 & 0.09 \plusmin 0.05 &-7.493&106.3&109.4& 1916 \\
EPIC 213546283& -0.17 \plusmin 0.06 & 0.18 \plusmin 0.02 &-43.53&105.1&103.6& 1619 & 0.545 \\
EPIC 210961508& 0.1 \plusmin 0.12 & 0.1 \plusmin 0.06 &-45.32&84.32&99.61& 1794 \\
EPIC 216111905& -0.23 \plusmin 0.12 & 0.17 \plusmin 0.07 &-82.87&47.34&98.95& 1293 \\
EPIC 212495601& -0.2 \plusmin 0.09 & 0.05 \plusmin 0.04 &-54.8&100&97.81& 1562 & 0.294\\
EPIC 201561956& -0.64 \plusmin 0.12 & 0.13 \plusmin 0.06 &-69.25&46.21&93.35& 1453 \\
TOI-832 & 0.4 \plusmin 0.08 & -0.02 \plusmin 0.04 &-78.89&14.48&82.26& 1403 \\
K2-248 & -0.13 \plusmin 0.08 & 0.1 \plusmin 0.04 &-31.36&63.91&81.33& 1748 \\
TOI-348 & 0.08 \plusmin 0.09 & -0.01 \plusmin 0.05 &-59.44&60.31&80.61& 1531 & 0.261 \\
TOI-844 & -0.03 \plusmin 0.08 & 0.03 \plusmin 0.04 &-55.98&60.92&80.3& 1622 & 0.18 \\
NGTS-4 & -0.17 \plusmin 0.06 & 0.22 \plusmin 0.03 &-77.27&82.75&73.46& 1435 & 103 \\
EPIC 211800191& -0.7 \plusmin 0.13 & 0.21 \plusmin 0.06 &-3.113&87.5&70.26& 2090 & 0.577 \\
K2-7 & -0.4 \plusmin 0.09 & 0.19 \plusmin 0.05 &-60.47&50.56&64.63& 1544 \\
TOI-924 & -0.28 \plusmin 0.05 & 0.1 \plusmin 0.02 &-52.45&87.46&56.11& 1624 & 0.367 \\
TOI-868 & -0.1 \plusmin 0.05 & 0.08 \plusmin 0.01 &-61.64&47&53.85& 1528 & 0.0956 \\
K2-73 & 0.02 \plusmin 0.05 & -0.02 \plusmin 0.02 &-22.03&72.03&51.98& 1825 & 0.0252 \\
TIC 287328202& -0.06 \plusmin 0.12 & 0.14 \plusmin 0.06 &-41.16&87.24&50.73& 1749 & 0.105 \\
HATS-52 & -0.31 \plusmin 0.11 & 0.06 \plusmin 0.05 &2.438&77.06&48.78& 2119 \\
EPIC 211736305 & 0.13 \plusmin 0.15 & 0.19 \plusmin 0.08 &-65.79&57.32&47.15& 1572 \\
K2-183 & 0.07 \plusmin 0.11 & 0.04 \plusmin 0.06 &-43.32&80.45&45.97& 1741 & 2.7 \\
EPIC 211978988 & 0.02 \plusmin 0.08 & -0.0 \plusmin 0.05 &-13.67&74.74&45.32& 2008 & 0.0737 \\
EPIC 212646483 & -0.27 \plusmin 0.08 & 0.13 \plusmin 0.04 &-68.48&51.5&43.55& 1348 \\
EPIC 218901589 & -0.12 \plusmin 0.05 & 0.06 \plusmin 0.02 &-31.92&69.22&43.24& 1748 & 0.033 \\
EPIC 220674823 & 0.09 \plusmin 0.09 & 0.08 \plusmin 0.05 &-44.9&56.29&43.21& 1681 & 0.0579 \\
TOI-810 & -0.24 \plusmin 0.05 & 0.12 \plusmin 0.02 &-38.64&64.44&11.85& 1709 & 2.53 \\

\end{tabular}
\end{center}
\end{table*}

\subsubsection{Chemical Abundance Relationships Between Short Period and Ultra-short Period Planets}\label{sec:popdiff}

There has been a great range of studies to link the chemical abundances of stars to the planets they host. The first link was discovering hot-Jupiters tending to favour iron-rich host stars \citep{hotjupfeh01,hotjupfeh05,hotjupfeh97}. \citet{Adibekyan12} showed that there is an overabundance of alpha-elements in short-period exoplanet hosts, in particular Neptune and super-Earth sized exoplanets, compared to stars hosting larger planets. \citet{Winn17} showed through iron abundances of planet hosts, that there was a population difference between hot-Jupiter's and their ultra short period counterparts, concluding that rocky USP planets might not necessarily be remnants of hot-Jovian cores. Further, \citet{Dai2019} also argued that short-period rocky worlds are more than likely  exposed rocky cores of sub-Neptunes, rather than hot-Jupiters. Per the discovery of the USP TOI-1444b, \citet{Dai2021} showed that hot-Neptunes tended to favour iron-rich stars, compared to their rockier counterparts. All of the above then motivates us, with our homogenous sample to explore the abundance trends between different short-period planet types.

Firstly, we split up our sample into five different categories; Ultra Hot Rocky (USR) exoplanets (R\textsubscript{P} < 2 \rearth; P\textsubscript{p} < 1 day; N = 9), Hot Rocky (HR) exoplanets (R\textsubscript{p} < 2 \rearth; 1 day $\leq$ P\textsubscript{p} < 10 days; N = 36), Ultra Hot Neptune (UHN) exoplanets  (2 \rearth $\leq$ R\textsubscript{p} < 4 \rearth; P\textsubscript{p} < 1 day; N = 3), Hot Neptune (HN) exoplanets  ( 2 \rearth $\leq$ R\textsubscript{p} < 4 \rearth; 1 day $\leq$ P\textsubscript{p} < 10 days; N = 48), and all other candidates that fit outside of these parameters. We plot the iron abundance, against the $\alpha$-element and magnesium abundances for these populations in Figure \ref{fig:alpha}. Within Figure \ref{fig:alpha}A, we see a range range of iron abundances for both HN and HR worlds. Near solar values, there is a similar distribution of these two populations, however beyond [Fe/H] > 0.15, there are a greater fraction of HNs compared to their rocky counterparts. We do see a difference in the magnesium abundances however between the two populations, as HR worlds tend to favour a wider range of magnesium abundances. In contrast, HNs are seen in more magnesium poor environments peaking at near Solar values. \citet{Adibekyan12} showed an over-abundance of magnesium for Neptune and super-Earth hosting stars, but these two populations were entangled within the same distribution for their comparison between hot-Jupiters, thus it is difficult to compare our results with theirs in this particular case. As with \citet{Dai2021}, we do see UHNs around stars enriched in iron compared to UHRs, with the one exception being the UHR exoplanet WASP-47 e, orbiting around an extremely iron-rich host ([Fe/H] = 0.45\plusmin0.09). There does also seem to be a trend in UHNs being preferentially found around higher [Mg/Fe] stars compared to UHRs, but this trend for ultra-short exoplanets severely weakens out for the $\alpha$-abundance (which is a combination of Ca, Mg, Ti and Al). With an exceedingly small sample size for both ultra-short period populations (N\textsubscript{UHR} = 9, N\textsubscript{UHN} = 3), more data and planets are needed to confirm the existence of such trends and their implications.

\begin{figure*}
	\includegraphics[width=\textwidth]{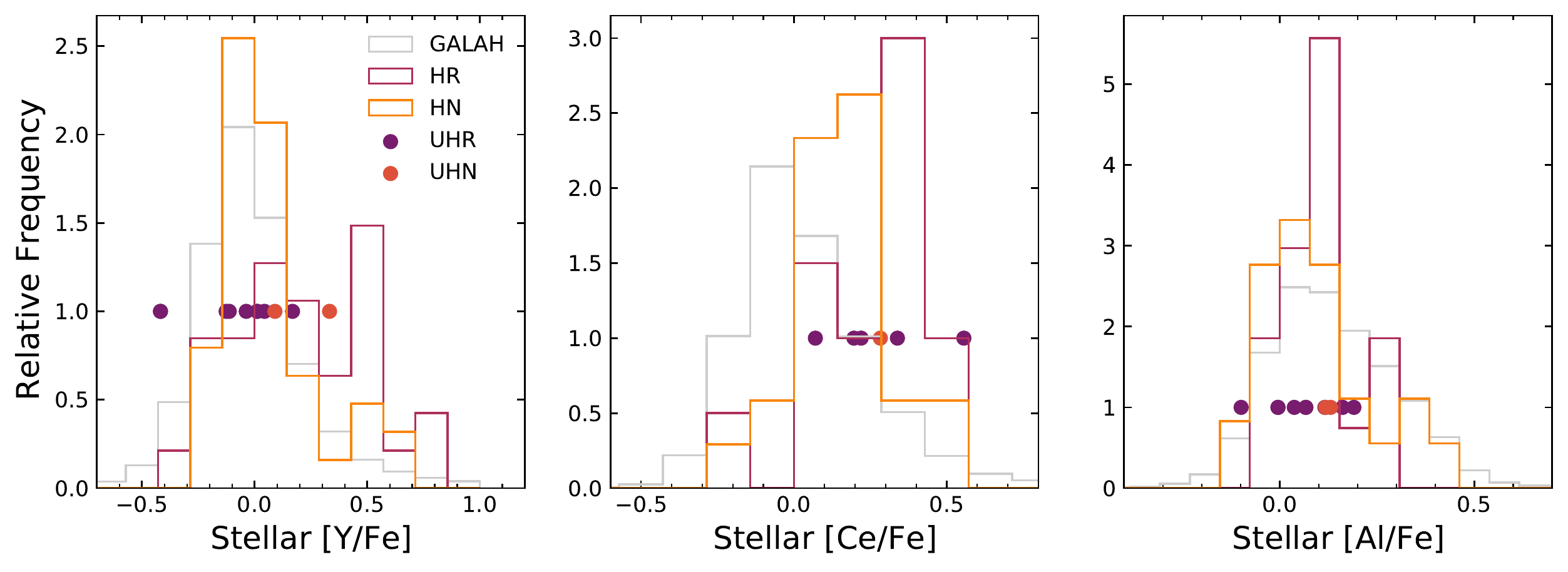}
 \caption{Chemical abundance distributions for the three elements, yttrium (Left), cerium (Middle) and aluminium (Right), that had p-values indicative of UN (pink) and HR (yellow) worlds being derived from different populations. We have included the overall GALAH abundance distributions (grey) along with the UHR (purple discs) and UHN (orange discs) for comparison as well.}\label{fig:YFe}
\end{figure*}

Similar to \cite{Winn17}, we wanted to see if there is a difference between the populations of HRs and HNs. In our case we not only have access to [Fe/H], but we also have access to over 20 abundances, which provides a more rigorous chemical test to see if our visual discrepancy in magnesium and $\alpha$ abundances between the two populations is real or not. We perform a two-sample Kolmogorov-Smirnov test between the HN and HR populations for all GALAH abundances. The Kolmogorov-Smirnov-statistic and p-value for each element is shown in Table \ref{tab:KStest}, sorted by their lowest p-value. Surprisingly, the  p-value between the two populations for the magnesium abundance was 0.052, meaning that their is no statistical evidence that the magnesium abundances for the two populations are significantly different from one another. Of the 29 abundances, only three distributions had p-values smaller than 0.05, those elements being Y, Ce, and Al. We have plotted the yttrium, cerium and aluminium abundances in Figure \ref{fig:YFe}.

The overall yttrium abundance distributions for HN and HR worlds is similar to the overall GALAH distribution, however we do see a trend where by more HR exoplanets are found around a broader range of [Y/Fe] values. We also see a higher fraction of HR planets orbiting stars enriched in yttrium ([Y/Fe] > 0.2) compared to their gaseous counterparts. As yttrium is known as a ``chemical-clock" element, whereby stars enriched in yttrium are typically younger, we see this trend as a possibility that HR planets tend to favour younger stars \citep{YMgGraph,YMgClockEvolved}. However, we also see a trend whereby HR worlds tend to favour magnesium-enriched stars. Magnesium is also a ``chemical-clock" element, but enhanced magnesium abundances are often a reflection of orbiting around older stars \citep{YMgGraph}. Thus, this positive correlation between magnesium and yttrium needs further investigation to determine as to why HR planets tend to favour stars enriched in both of these elements. Cerium has similar trends, with HR stars being enriched with Ce compared to HN hosting stars. However, both HN and HR [Ce/Fe] abundances peak super-Solar, compared to the Solar-like abundance peak for the overall GALAH sample, with the interpretation to this distribution remaining an interesting development for future work.  

Overall, there does not seem to be a statistical difference between the chemical properties of stars that host HNs and those that host HRs. Thus, following the conclusions of \citet{Dai2021} and \cite{Winn17}, our more robust chemical abundance results show that there is a possibility that short-period rocky worlds might be the remnant cores of hotter-gaseous Neptune worlds. The reason being is that there are only three of the 29 elemental abundances that had KS statistics significant enough to show that HRs and HNs come from different populations. Even with these three elements, the smallest p-value of 0.01, along with a relatively small sample size of our HRs and HNs, there is still more research to be done to determine the similarities in these populations, and thus determine the origins and evolution of short-period exoplanets.

\begin{table}
\caption{Our two-sample Kolmogorov-Smirnov test results between our Hot Neptune and Hot Rocky planet samples for each GALAH abundance.}\label{tab:KStest}
\begin{center}
\begin{tabular}{ccc}
\hline
Element & D-statistic & p-value \\
\hline
    Y & 0.364 & 0.01 \\
    Ce & 0.476 & 0.024 \\
    Al & 0.303 & 0.037 \\
    Mg & 0.299 & 0.052 \\
    Na & 0.25 & 0.132 \\
    Cr & 0.238 & 0.169 \\
    Ba & 0.236 & 0.176 \\
    $\alpha$ & 0.217 & 0.254 \\
    Cu & 0.212 & 0.275 \\
    K & 0.21 & 0.302 \\
    Mo & 0.625 & 0.303 \\
    Zr & 0.326 & 0.311 \\
    C & 0.571 & 0.318 \\
    O & 0.235 & 0.323 \\
    La & 0.367 & 0.357 \\
    Si & 0.2 & 0.357 \\
    Mn & 0.19 & 0.39 \\
    Rb & 0.889 & 0.4 \\
    Sm & 0.667 & 0.4 \\
    Sc & 0.183 & 0.46 \\
    Li & 0.333 & 0.5 \\
    Ca & 0.166 & 0.571 \\
    Co & 0.201 & 0.573 \\
    Fe & 0.16 & 0.622 \\    
    Zn & 0.16 & 0.629 \\
    Ti & 0.147 & 0.707 \\
    V & 0.159 & 0.821 \\
    Ni & 0.11 & 0.945 \\
    Nd & 0.667 & 1.0 \\
\end{tabular}
\end{center}
\end{table}

\subsection{Assessing the radius valley and super-Earth desert}

\begin{figure*}
 \centering
	\includegraphics[width=\columnwidth]{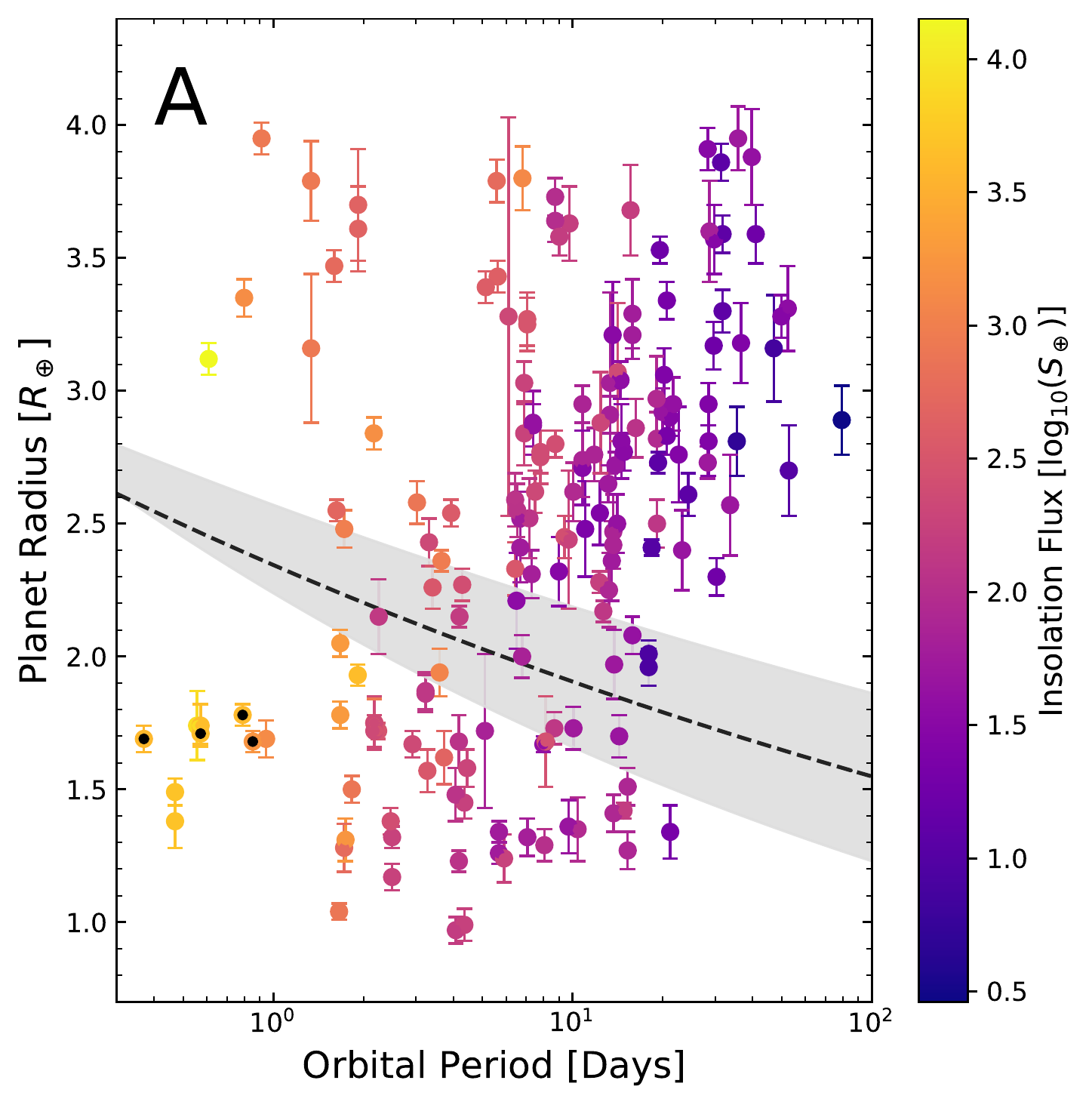}
    \includegraphics[width=\columnwidth]{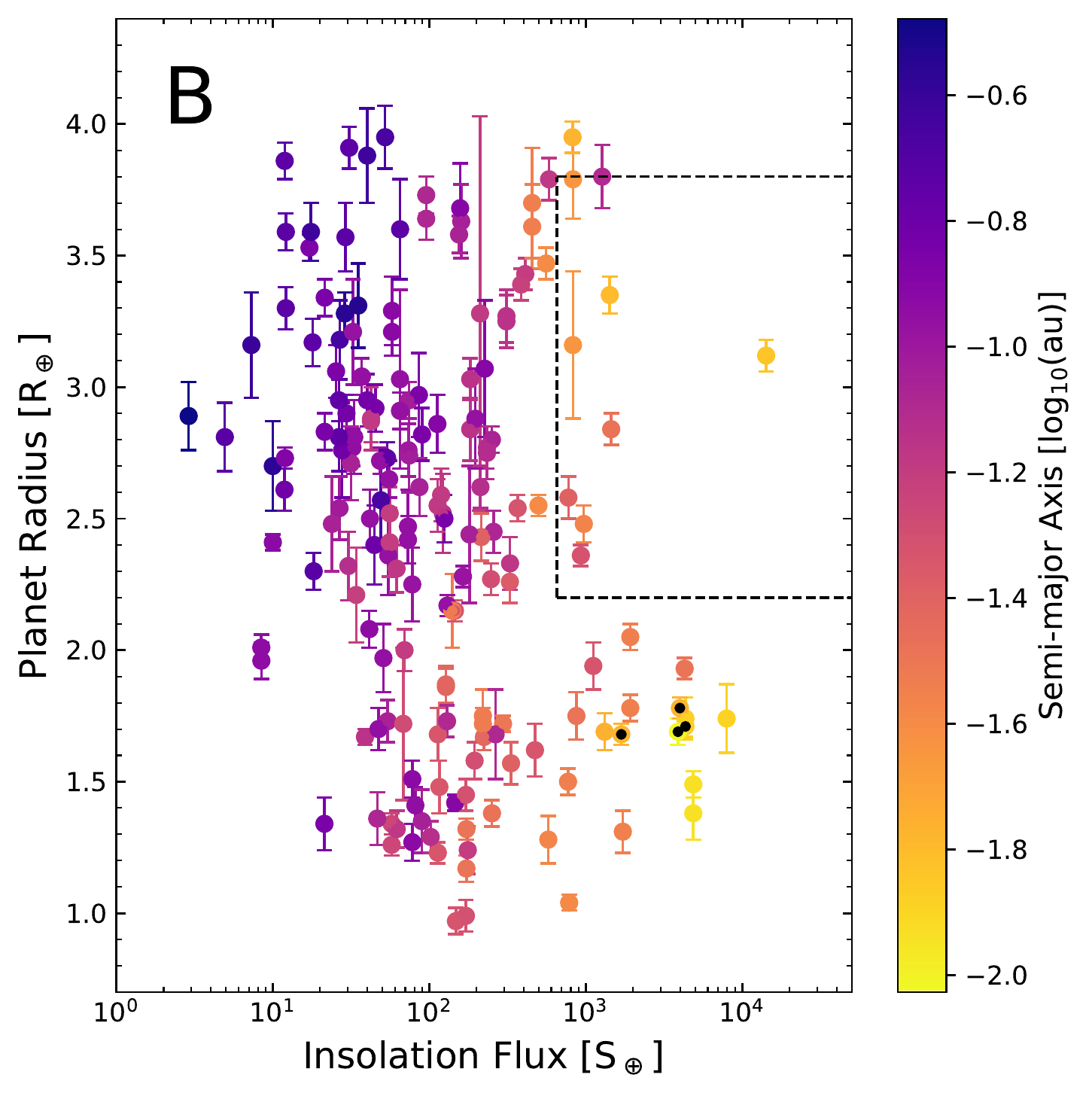}
 \caption{Left: Our confirmed and candidate exoplanet planetary radius values are plotted against their orbital periods, each being coloured by their insolation flux values. The dashed line and shaded region around it, indicates the slope and error in the radius valley as noted by \citet{evapSlope} Right: Planet and candidate radius values are plotted against their incident flux, coloured by their semi-major axis values. The dashed lines enclose the hot Super-Earth desert noted by \citet{superDesertBox}. Our four ultra-short period super-Earths are depicted in each plot by black dots.}\label{fig:evap}
\end{figure*}

Because we have four ultra-short period exoplanets within our sample straddling the super-Earth radius gap, we show in Figure \ref{fig:evap}a their location in the so-called two-dimensional radius gap, i.e. the planetary radius compared to the orbital period \citep{FultonLumps,evapSlope}. None of the ultra-short period super-Earths within our sample lie within the photoevaporation valley, the grey dashed line within the Figure \ref{fig:evap}a, all below this line. Nor do any of these ultra-short period have periods smaller than the expected Roche limits of their host stars \citep{Rappaport2013}. They do however has orbits that are smaller than the dust sublimation radius of their host stars \citep{Isella06}, meaning that in-situ formation of these exoplanets is unlikely. There is only one confirmed-exoplanet that lies within 1-sigma of the line proposed by \citet{evapSlope}, that being K2-247 b (P: 2.25~d, R\textsubscript{p}: 2.15\plusmin0.14~\rearth) \citep{Livingston18}.

We show also in Figure \ref{fig:evap} the insolation flux of confirmed and candidate exoplanets compared to their estimated radii; the dashed lines enclose a region of parameter space known as the Neptunian desert, proposed by \citep{superDesertBox}. This Neptunian desert is a region of flux-radius phase space where super-Earths are less common, explained by evaporation of volatile elements. The sub-Neptune NGTS-4 b (S\textsubscript{eff}: 824\plusmin62~S\textsubscript{$\oplus$}, R\textsubscript{p}: 3.16\plusmin0.29\rearth) is the only confirmed exoplanet within this region \citep{NGTS4b} with two TOIs, 1926.01 and 1948.01, also being contained within the Neptunian desert.

As mentioned previously, the origin of ultra-short period super-Earths remains an active field of science \citep{Wagner12,superDesertBox,supEarthAtmos,Spaargaren20}. The elemental abundances of refractory elements, such as Mg, Si, Fe and molar ratios of such elements can help constrain the interiors of rocky worlds \citep{Unterborn2016,Seager2007,Valencia2006}. Modelling the interiors of ultra-short period can help determine if these worlds are likely remnants of Jovian cores, or super-Earths that have migrated inwards to their current positions \citep{Mocquet2014,mercuryorigin}. 44\% of ultra-short period super-Earths discovered are enriched with iron, with exoplanet candidate KOI 1843.03 predicted to have a core-mass fraction as high as mercury's. However, there seems to be no trend between a rocky planet's core-mass fraction and its received flux values \citep{Prince2020}. Mg/Si and Fe/Si ratios from GALAH DR3 could be used to further constrain the chemical and geological composition of these super-Earths using models such as \citet{Dorn2015,Dorn2017,EarthWang} and \citet{exoplex}. But this is outside the scope of this paper and is left for further future investigation.

\section{Conclusion}

The wealth of astronomical data in large galactic archaeology surveys like GALAH can be used in numerous astrophysical fields, including exoplanetary science. We have cross-matched GALAH's latest release with the latest astrometric data from \textit{Gaia} EDR3 to determine the physical parameters of 279 confirmed and candidate exoplanet hosting stars. With these new stellar parameters and combining planetary observables from transit photometry and Doppler spectroscopy where applicable, we have determined and refined the physical characteristics of 105 confirmed exoplanets, 146 K2 candidates, 95 TOIs and 52 CTOIs, along with analysing the chemical abundances of these host stars. From our study we have uncovered:

\begin{itemize}
    \item 30 CTOI and K2 candidates have radii larger than our planetary limit of R\textsubscript{P} $<$ 2 R\textsubscript{J}, meaning that these candidates would be more akin to brown-dwarfs or stellar companions. TOIs 147.01, 565.01, 959.01, 1072.01 and 2391.01 are also too large to be planetary in nature. \\
    \item For the majority of our candidates, their host star's \vsini values are above 5\kms, with predicted RV signals being less than 5\ms. We then predict that it will be difficult, with current methodology, to obtain mass confirmations for most of the smaller planet candidates orbiting stars found within our sample. In the extreme cases for smaller planets orbiting rapid-rotating stars it will be highly improbable to derive their mass measurements, with TOI-1219.01 needing over 14,000 RV measurements to confirm its mass from a typical modern spectrograph.\\
    \item Out of the homogeneous data set of 105 exoplanets with new planetary parameters, we have updated parameters for five ultra-short period super-Earths -- K2-106 b, K2-216 b, WASP-47 e, K2-131 b and CoRoT- 7 b -- which make up 28\% of all such planets with known mass and radius values. In particular, our refined radius and mass values for WASP-47 e, K2-106 b and CoRoT-7 b have uncertainties smaller than 2.3\% and 8.5\%, respectively. From these refinements, K2-106 b's equilibrium temperature of 2570 K far exceeds the condensation temperature of most refractory elements, making it the hottest super-Earth to date. With our mass and radius measurements of CoRoT-7 b, 1.68\plusmin0.04\rearth~and 4.94\plusmin0.41\mearth~respectively, its radius is now large enough for an atmosphere to contribute to its overall radius and now straddles the super-Earth radius gap. \\
    \item Through stellar chemo-kinematic and dynamical data, we announce that three confirmed planet hosts, including NGTS-4, K2-183 and K2-337 along with candidate host stars TOI-810 and EPIC 211064647 to be members of the Milky Way's thick disc. By knowing more thick-disc hosts, we will be able to better determine the statistics of planetary architectures across stellar populations and determine the habitability of systems across the galaxy.\\
    \item With GALAH chemical abundances, we have shown that there does not seem to be a statistical difference between the chemical properties of stars that host Hot Neptunes and those that host Hot Rocky exoplanets. Thus, there is a possibility that short-period rocky worlds might be the remnant cores of hotter-gaseous Neptune-sized worlds.
\end{itemize}

\section*{Software}
\textsc{AstroPy} \citep{astropy}, \textsc{Astroquery} \citep{astroquery}, \textsc{Isochrones} \citep{isochrones}, \textsc{Matplotlib} \citep{matplotlib}, \textsc{MultiNest} \citep{multinest19,multinest09,multinest08}, \textsc{Multiprocessing} \citep{multiprocessing}, \textsc{NumPy} \citep{numpy1,numpy2}, \textsc{Pandas} \citep{pandas}, \textsc{SciPy} \citep{scipy}

\section*{Acknowledgements}

We thank the Australian Time Allocation Committee for their generous allocation of AAT time, which made this work possible. Our research is based upon data acquired through the Australian Astronomical Observatory. We acknowledge the traditional owners of the land on which the AAT stands, the Gamilaraay people, and pay our respects to elders past, present and emerging. 

This research has made use of the NASA Exoplanet Archive, which is operated by the California Institute of Technology, under contract with the National Aeronautics and Space Administration under the Exoplanet Exploration Program.

This work has made use of the TIC and CT Stellar Properties Catalog, through the \textit{TESS} Science Office's target selection working group (architects K. Stassun, J. Pepper, N. De Lee, M. Paegert, R. Oelkers). The Filtergraph data portal system is trademarked by Vanderbilt University.

This research has made use of the Exoplanet Follow-up Observation Program website, which is operated by the California Institute of Technology, under contract with the National Aeronautics and Space Administration under the Exoplanet Exploration Program.

This work has made use of data from the European Space Agency (ESA) mission \textit{Gaia} (\url{http://www.cosmos.esa.int/gaia}), processed by the \textit{Gaia} Data Processing and Analysis Consortium (DPAC, \url{http://www.cosmos.esa.int/web/gaia/dpac/}
consortium). Funding for the DPAC has been provided by national institutions, in particular the institutions participating in the \textit{Gaia} Multilateral Agreement.

J.T.C would like to thank BC and DN, and is supported by the Australian Government Research Training Program (RTP) Scholarship.

\section*{Data Availability}
The data underlying this article are available in the article and in its online supplementary material.




\bibliographystyle{mnras}
\bibliography{example} 




\appendix

\section{Data Tables}

\begin{landscape}
\begin{table}
\caption{Our stellar parameters for confirmed and candidate exoplanet hosting stars. The full table of which can be found on the online version of this paper.}
\label{tab:stellarparams}
\begin{tabular}{lllllllllll}

\hline
TIC ID & 2MASS ID & GAIA DR3 ID & RA & Dec & T\textsubscript{eff} & log\,$g$ & [Fe/H] & [$\alpha$/Fe] & [M/H] & \nodata \\
 & & & [deg] & [deg] & [K] & [cgs] & [dex] & [dex] & [dex] & \nodata \\
\hline
355703913 & 00030587-6228096 & 4904523077718046720 & 0.7745$\pm$0.0099 & -62.4693$\pm$0.0096 & 5451$\pm$112 & 4.50$\pm$0.19 & 0.23$\pm$0.10 & 0.05$\pm$0.06 & 0.30$\pm$0.14 & \nodata \\
70785910 & 00062006-3020105 & 2320600658577841536 & 1.5836$\pm$0.0152 & -30.3364$\pm$0.0115 & 5819$\pm$96 & 4.09$\pm$0.19 & 0.08$\pm$0.09 & -0.01$\pm$0.05 & 0.06$\pm$0.12 & \nodata \\
201256771 & 00071640-5827263 & 4918333184280735232 & 1.8187$\pm$0.0083 & -58.4573$\pm$0.0084 & 4960$\pm$81 & 3.42$\pm$0.18 & -0.34$\pm$0.05 & 0.19$\pm$0.02 & -0.04$\pm$0.06 & \nodata \\
328066864 & 00392335+0443037 & 2554244130700741504 & 9.8475$\pm$0.0148 & 4.7178$\pm$0.0077 & 4742$\pm$123 & 4.59$\pm$0.19 & -0.12$\pm$0.12 & 0.06$\pm$0.08 & -0.04$\pm$0.17 & \nodata \\
299897516 & 00403884-6052497 & 4903399170675741568 & 10.1621$\pm$0.0099 & -60.8804$\pm$0.0092 & 5417$\pm$98 & 3.86$\pm$0.18 & -0.83$\pm$0.09 & 0.26$\pm$0.05 & -0.42$\pm$0.12 & \nodata \\
418761354 & 00455526+0620490 & 2556231154370582400 & 11.4802$\pm$0.0200 & 6.3470$\pm$0.0111 & 4602$\pm$88 & 4.61$\pm$0.18 & -0.04$\pm$0.07 & 0.03$\pm$0.03 & 0.01$\pm$0.09 & \nodata \\
234504626 & 00474568-6225232 & 4902237021244807296 & 11.9408$\pm$0.0091 & -62.4232$\pm$0.0091 & 5001$\pm$90 & 4.55$\pm$0.18 & 0.18$\pm$0.07 & 0.03$\pm$0.03 & 0.23$\pm$0.09 & \nodata \\
257394133 & 00485666+0524595 & 2552929320952237952 & 12.2361$\pm$0.0169 & 5.4164$\pm$0.0086 & 5617$\pm$96 & 4.32$\pm$0.19 & 0.27$\pm$0.08 & -0.02$\pm$0.04 & 0.25$\pm$0.10 & \nodata \\
257218673 & 00490407+0215596 & 2549922465888108032 & 12.2670$\pm$0.0146 & 2.2666$\pm$0.0080 & 5579$\pm$93 & 4.06$\pm$0.18 & 0.21$\pm$0.08 & 0.04$\pm$0.04 & 0.27$\pm$0.10 & \nodata \\
405336406 & 00493540+1001123 & 2582183683235097728 & 12.3975$\pm$0.0185 & 10.0200$\pm$0.0142 & 4490$\pm$76 & 1.67$\pm$0.25 & -0.74$\pm$0.05 & 0.20$\pm$0.01 & -0.43$\pm$0.05 & \nodata \\
266012991 & 00510476+0931003 & 2581918597853527424 & 12.7700$\pm$0.0204 & 9.5168$\pm$0.0144 & 5554$\pm$77 & 4.38$\pm$0.18 & -0.02$\pm$0.05 & 0.04$\pm$0.02 & 0.04$\pm$0.06 & \nodata \\
333605244 & 00510570-0111452 & 2535901287452166400 & 12.7737$\pm$0.0167 & -1.1959$\pm$0.0089 & 4975$\pm$125 & 4.59$\pm$0.19 & 0.04$\pm$0.11 & 0.07$\pm$0.06 & 0.14$\pm$0.14 & \nodata \\
257434940 & 00511854+0520004 & 2552891834477747200 & 12.8273$\pm$0.0152 & 5.3334$\pm$0.0093 & 5254$\pm$98 & 3.91$\pm$0.19 & -0.13$\pm$0.08 & 0.13$\pm$0.05 & 0.06$\pm$0.11 & \nodata \\
266015990 & 00521914+1047409 & 2582617711154563968 & 13.0800$\pm$0.0146 & 10.7947$\pm$0.0105 & 5499$\pm$104 & 4.40$\pm$0.19 & 0.09$\pm$0.09 & 0.08$\pm$0.05 & 0.21$\pm$0.12 & \nodata \\
257435562 & 00523368+0330277 & 2551651031310145024 & 13.1404$\pm$0.0162 & 3.5078$\pm$0.0090 & 5750$\pm$88 & 4.21$\pm$0.18 & 0.10$\pm$0.07 & 0.05$\pm$0.03 & 0.18$\pm$0.09 & \nodata \\
266017624 & 00524666+0941345 & 2581916673708201216 & 13.1944$\pm$0.0163 & 9.6929$\pm$0.0129 & 5014$\pm$91 & 4.54$\pm$0.19 & 0.33$\pm$0.07 & -0.01$\pm$0.04 & 0.31$\pm$0.09 & \nodata \\
344586726 & 00530642+0606009 & 2553369877222007168 & 13.2768$\pm$0.0152 & 6.1002$\pm$0.0108 & 4935$\pm$116 & 4.61$\pm$0.19 & -0.05$\pm$0.10 & -0.02$\pm$0.06 & -0.07$\pm$0.14 & \nodata \\
266030611 & 00534179+1002417 & 2581964605543263104 & 13.4241$\pm$0.0130 & 10.0448$\pm$0.0107 & 5652$\pm$88 & 4.44$\pm$0.18 & 0.35$\pm$0.07 & -0.00$\pm$0.03 & 0.35$\pm$0.09 & \nodata \\
\nodata & \nodata & \nodata & \nodata & \nodata & \nodata & \nodata & \nodata & \nodata & \nodata & \nodata \\
\hline
\\[2ex]
\hline
R\textsubscript{s} & M\textsubscript{s} & L\textsubscript{s} & Age & RV & vsini & HZ\textsubscript{1} & HZ\textsubscript{2} & HZ\textsubscript{3} & HZ\textsubscript{4} & HZ\textsubscript{5} \\
{[}R\textsubscript{$\odot$}] & [\msun] & [L\textsubscript{$\odot$}] & [log\textsubscript{10}(Gyr)] & [kms\textsuperscript{-1}] & [kms\textsuperscript{-1}] & [au] & [au] & [au] & [au] & [au] \\
\hline
0.925$\pm$0.015 & 0.977$\pm$0.040 & 0.677$\pm$0.060 & 9.526$\pm$0.424 & 17.286$\pm$0.096 & 5.168$\pm$2.488 & 0.626 & 0.823 & 0.828 & 1.429 & 1.495 \\
1.610$\pm$0.033 & 1.106$\pm$0.065 & 2.660$\pm$0.208 & 9.851$\pm$0.130 & 31.222$\pm$0.123 & 5.488$\pm$2.350 & 1.222 & 1.597 & 1.617 & 2.745 & 2.871 \\
3.233$\pm$0.059 & 1.220$\pm$0.098 & 5.664$\pm$0.425 & 9.693$\pm$0.130 & -11.577$\pm$0.071 & 6.509$\pm$2.133 & 1.845 & 2.442 & 2.441 & 4.314 & 4.513 \\
0.727$\pm$0.011 & 0.761$\pm$0.025 & 0.240$\pm$0.026 & 9.805$\pm$0.398 & -2.143$\pm$0.081 & 6.950$\pm$2.492 & 0.382 & 0.507 & 0.506 & 0.904 & 0.946 \\
1.765$\pm$0.036 & 0.928$\pm$0.037 & 2.400$\pm$0.200 & 10.041$\pm$0.056 & 1.605$\pm$0.117 & 9.993$\pm$2.320 & 1.180 & 1.553 & 1.561 & 2.699 & 2.823 \\
0.688$\pm$0.010 & 0.723$\pm$0.019 & 0.190$\pm$0.016 & 9.719$\pm$0.419 & -26.481$\pm$0.081 & 6.666$\pm$2.247 & 0.342 & 0.454 & 0.453 & 0.814 & 0.853 \\
0.801$\pm$0.011 & 0.865$\pm$0.024 & 0.359$\pm$0.028 & 9.533$\pm$0.425 & 20.991$\pm$0.073 & 6.239$\pm$2.250 & 0.464 & 0.614 & 0.614 & 1.082 & 1.132 \\
1.167$\pm$0.027 & 1.019$\pm$0.042 & 1.214$\pm$0.100 & 9.912$\pm$0.128 & -4.166$\pm$0.104 & 6.379$\pm$2.301 & 0.832 & 1.092 & 1.101 & 1.887 & 1.973 \\
1.602$\pm$0.035 & 1.078$\pm$0.054 & 2.228$\pm$0.178 & 9.952$\pm$0.188 & 4.115$\pm$0.114 & 7.170$\pm$2.284 & 1.129 & 1.482 & 1.494 & 2.564 & 2.681 \\
24.617$\pm$1.181 & 1.212$\pm$0.296 & 220.397$\pm$25.991 & 9.593$\pm$0.280 & -148.747$\pm$0.085 & 7.090$\pm$2.096 & 11.700 & 15.531 & 15.481 & 27.974 & 29.309 \\
1.010$\pm$0.017 & 0.929$\pm$0.029 & 0.868$\pm$0.057 & 9.951$\pm$0.124 & 13.396$\pm$0.082 & 4.899$\pm$2.143 & 0.706 & 0.927 & 0.934 & 1.604 & 1.678 \\
0.767$\pm$0.011 & 0.825$\pm$0.025 & 0.323$\pm$0.034 & 9.522$\pm$0.449 & 18.833$\pm$0.104 & 3.666$\pm$2.651 & 0.440 & 0.583 & 0.582 & 1.028 & 1.076 \\
1.794$\pm$0.039 & 1.003$\pm$0.035 & 2.195$\pm$0.190 & 10.033$\pm$0.081 & 11.729$\pm$0.115 & 10.051$\pm$2.262 & 1.135 & 1.498 & 1.502 & 2.618 & 2.738 \\
0.969$\pm$0.017 & 0.969$\pm$0.045 & 0.769$\pm$0.064 & 9.756$\pm$0.321 & -16.011$\pm$0.096 & 6.027$\pm$2.367 & 0.665 & 0.875 & 0.880 & 1.517 & 1.586 \\
1.337$\pm$0.024 & 1.086$\pm$0.043 & 1.749$\pm$0.124 & 9.837$\pm$0.129 & -6.373$\pm$0.109 & 5.812$\pm$2.260 & 0.994 & 1.300 & 1.315 & 2.239 & 2.342 \\
0.817$\pm$0.013 & 0.871$\pm$0.028 & 0.378$\pm$0.030 & 9.677$\pm$0.407 & -5.682$\pm$0.068 & 6.797$\pm$2.251 & 0.475 & 0.629 & 0.629 & 1.109 & 1.160 \\
0.728$\pm$0.012 & 0.773$\pm$0.025 & 0.282$\pm$0.028 & 9.655$\pm$0.444 & 8.326$\pm$0.083 & 6.188$\pm$2.447 & 0.412 & 0.545 & 0.545 & 0.964 & 1.009 \\
1.037$\pm$0.019 & 1.062$\pm$0.042 & 0.982$\pm$0.071 & 9.527$\pm$0.378 & -9.459$\pm$0.098 & 7.432$\pm$2.237 & 0.747 & 0.980 & 0.989 & 1.692 & 1.769 \\
\nodata & \nodata & \nodata & \nodata & \nodata & \nodata & \nodata & \nodata & \nodata & \nodata & \nodata \\
\hline

\end{tabular}
\end{table}
\end{landscape}

\begin{landscape}
\begin{table}
\caption{Our refined confirmed and candidate exoplanet parameters. The full table of which can be found on the online version of this paper.}
\label{tab:planetparams}
\begin{tabular}{lllllllllllll}
 
\hline
TIC ID & Candidate & Period & Semi-major axis & R\textsubscript{p}/R\textsubscript{*} & au/R\textsubscript{*} & Transit Depth & Inclination & Eccentricity & K\textsubscript{amp} & \nodata \\
& Name & [Days] & [au] & & & [\%] & [deg] & & [ms\textsuperscript{-2}] & \nodata \\
\hline
677945 & K2-128 b & 5.675729$\pm$0.0001680 & 0.0559$\pm$0.0004 & 0.01709$\pm$0.00059 & 17.8385$\pm$0.268 & 0.029239$\pm$0.001693 & 88.331$\pm$0.574 & 0.23$\pm$0.221 & \nodata & \nodata \\
677945 & EPIC 212686205.01 & 5.67556$\pm$0.0002900 & 0.0559$\pm$0.0004 & 0.01822$\pm$0.00058 & 17.8382$\pm$0.268 & 0.0332$\pm$0.001700 & \nodata & \nodata & \nodata & \nodata \\
678699 & EPIC 212560683.01 & 13.7043$\pm$0.0037000 & 0.1134$\pm$0.0012 & 0.01257$\pm$0.0006 & 23.7383$\pm$0.463 & 0.0158$\pm$0.001300 & \nodata & \nodata & \nodata & \nodata \\
706595 & K2-194 b & 39.72342$\pm$0.0031880 & 0.2358$\pm$0.0034 & 0.02562$\pm$0.00128 & 36.5615$\pm$0.898 & 0.06572$\pm$0.005381 & 89.065$\pm$0.46 & \nodata & \nodata & \nodata \\
706595 & EPIC 212672300.01 & 39.703624$\pm$0.0021210 & 0.2358$\pm$0.0034 & 0.02721$\pm$0.00093 & 36.5523$\pm$0.897 & 0.0741$\pm$0.002800 & \nodata & \nodata & \nodata & \nodata \\
707724 & EPIC 212585579.01 & 3.021889$\pm$0.0000460 & 0.0411$\pm$0.0007 & 0.03443$\pm$0.00114 & 7.8281$\pm$0.188 & 0.1186$\pm$0.004500 & \nodata & \nodata & \nodata & \nodata \\
743782 & EPIC 212652418.01 & 19.1324$\pm$0.0031010 & 0.1416$\pm$0.0024 & 0.01912$\pm$0.00075 & 22.5198$\pm$0.579 & 0.0366$\pm$0.002000 & \nodata & \nodata & \nodata & \nodata \\
771548 & EPIC 212587672.01 & 23.231764$\pm$0.0008410 & 0.1579$\pm$0.0025 & 0.02175$\pm$0.00135 & 33.5908$\pm$0.802 & 0.047326$\pm$0.005322 & 89.421$\pm$0.422 & \nodata & \nodata & \nodata \\
771548 & EPIC 212587672.02 & 15.2841$\pm$0.0020000 & 0.1195$\pm$0.0019 & 0.01367$\pm$0.00066 & 25.4093$\pm$0.607 & 0.0187$\pm$0.001500 & \nodata & \nodata & \nodata & \nodata \\
771548 & EPIC 212587672 b & 15.2841$\pm$0.0017600 & 0.1195$\pm$0.0019 & 0.01152$\pm$0.00065 & 25.4108$\pm$0.606 & 0.01328$\pm$0.001324 & \nodata & \nodata & \nodata & \nodata \\
1103432 & 565.01 & 3.727891$\pm$0.0004550 & 0.0496$\pm$0.001 & 0.13363$\pm$0.00451 & 6.2069$\pm$0.178 & 19.571261$\pm$0.548862 & \nodata & \nodata & \nodata & \nodata \\
2621212 & K2-199 c & 7.374442$\pm$0.0000660 & 0.0663$\pm$0.0006 & 0.03868$\pm$0.0012 & 20.9489$\pm$0.324 & 0.149668$\pm$0.007299 & 88.893$\pm$0.557 & \nodata & \nodata & \nodata \\
2621212 & K2-199 b & 3.225392$\pm$0.0000320 & 0.0382$\pm$0.0004 & 0.02522$\pm$0.00089 & 12.0707$\pm$0.187 & 0.063616$\pm$0.003787 & 87.182$\pm$1.475 & \nodata & \nodata & \nodata \\
2621212 & EPIC 212779596.01 & 3.225349$\pm$0.0000430 & 0.0382$\pm$0.0004 & 0.02506$\pm$0.001 & 12.0704$\pm$0.187 & 0.062827$\pm$0.004389 & \nodata & \nodata & \nodata & \nodata \\
2621212 & EPIC 212779596.02 & 7.374284$\pm$0.0001100 & 0.0663$\pm$0.0006 & 0.03876$\pm$0.00161 & 20.9486$\pm$0.324 & 0.150314$\pm$0.011041 & \nodata & \nodata & \nodata & \nodata \\
2670610 & EPIC 213546283.01 & 9.770225$\pm$0.0001750 & 0.0898$\pm$0.0012 & 0.02896$\pm$0.00123 & 16.8145$\pm$0.375 & 0.083943$\pm$0.005663 & 88.885$\pm$0.815 & \nodata & \nodata & \nodata \\
2712931 & EPIC 212645891.01 & 0.328227$\pm$0.0000010 & 0.0092$\pm$0.0002 & 0.04742$\pm$0.00135 & 1.7434$\pm$0.048 & 0.224802$\pm$0.001692 & 41.646$\pm$5.432 & \nodata & \nodata & \nodata \\
2764004 & EPIC 212624936.01 & 11.81387$\pm$0.0008500 & 0.0994$\pm$0.0015 & 0.02686$\pm$0.00108 & 22.7423$\pm$0.555 & 0.0722$\pm$0.004200 & \nodata & \nodata & \nodata & \nodata \\
\nodata & \nodata & \nodata & \nodata & \nodata & \nodata & \nodata & \nodata & \nodata & \nodata & \nodata \\
\hline
\\[2ex]
\hline
R\textsubscript{p} & M\textsubscript{p} & Flag\textsubscript{mass} & $\rho$\textsubscript{p} & S\textsubscript{eff} & T\textsubscript{eq WM} & T\textsubscript{eq HD} & g\textsubscript{p} & $v_e$ & Flag\textsubscript{cat}\\
\rearth & \mearth & & gcm\textsuperscript{-3} & S\textsubscript{$\oplus$} & [K] & [K] & [ms\textsuperscript{-2}] & [ms\textsuperscript{-1}] & \\
\hline
1.26$\pm$0.04 & \nodata & \nodata & \nodata & 57.26$\pm$4.49 & 732$\pm$19 & 871$\pm$23 & \nodata & \nodata & EXO \\
1.34$\pm$0.04 & \nodata & \nodata & \nodata & 57.26$\pm$4.48 & 732$\pm$19 & 871$\pm$23 & \nodata & \nodata & K2 \\
1.41$\pm$0.07 & \nodata & \nodata & \nodata & 81.14$\pm$5.28 & 799$\pm$19 & 950$\pm$23 & \nodata & \nodata & K2 \\
3.88$\pm$0.18 & \nodata & \nodata & \nodata & 39.93$\pm$3.14 & 687$\pm$15 & 817$\pm$18 & \nodata & \nodata & EXO \\
4.12$\pm$0.12 & \nodata & \nodata & \nodata & 39.95$\pm$3.14 & 630$\pm$30 & 749$\pm$35 & \nodata & \nodata & K2 \\
4.24$\pm$0.12 & \nodata & \nodata & \nodata & 797.66$\pm$62.96 & 1331$\pm$63 & 1583$\pm$75 & \nodata & \nodata & K2 \\
2.82$\pm$0.1 & \nodata & \nodata & \nodata & 89.55$\pm$7.88 & 840$\pm$20 & 999$\pm$24 & \nodata & \nodata & K2 \\
2.4$\pm$0.15 & \nodata & \nodata & \nodata & 44.41$\pm$3.51 & 705$\pm$16 & 839$\pm$19 & \nodata & \nodata & K2 \\
1.51$\pm$0.07 & \nodata & \nodata & \nodata & 77.63$\pm$6.14 & 790$\pm$21 & 940$\pm$25 & \nodata & \nodata & K2 \\
1.27$\pm$0.07 & \nodata & \nodata & \nodata & 77.63$\pm$6.14 & 790$\pm$21 & 940$\pm$25 & \nodata & \nodata & EXO \\
25.07$\pm$0.65 & \nodata & \nodata & \nodata & 1429.21$\pm$119.76 & 1420$\pm$84 & 1689$\pm$99 & \nodata & \nodata & TOI \\
2.87$\pm$0.08 & \nodata & \nodata & \nodata & 42.3$\pm$3.22 & 697$\pm$15 & 828$\pm$18 & \nodata & \nodata & EXO \\
1.87$\pm$0.07 & \nodata & \nodata & \nodata & 127.37$\pm$9.71 & 894$\pm$23 & 1063$\pm$28 & \nodata & \nodata & EXO \\
1.86$\pm$0.07 & \nodata & \nodata & \nodata & 127.39$\pm$9.71 & 894$\pm$23 & 1064$\pm$28 & \nodata & \nodata & K2 \\
2.88$\pm$0.12 & \nodata & \nodata & \nodata & 42.29$\pm$3.22 & 697$\pm$15 & 828$\pm$18 & \nodata & \nodata & K2 \\
3.63$\pm$0.14 & \nodata & \nodata & \nodata & 159.39$\pm$11.64 & 971$\pm$20 & 1154$\pm$24 & \nodata & \nodata & K2 \\
5.87$\pm$0.12 & \nodata & \nodata & \nodata & 13416.93$\pm$1670.66 & 2692$\pm$143 & 3201$\pm$171 & \nodata & \nodata & K2 \\
2.75$\pm$0.1 & \nodata & \nodata & \nodata & 73.34$\pm$6.44 & 799$\pm$19 & 951$\pm$23 & \nodata & \nodata & K2 \\
\nodata & \nodata & \nodata & \nodata & \nodata & \nodata & \nodata & \nodata & \nodata & \nodata \\ 
\hline

\end{tabular}
\end{table}
\end{landscape}




\bsp	
\label{lastpage}
\end{document}